\newcommand{\mpt}{{/\!\!\!\! \vec{P}_T}} 

\documentclass[twocolumn,aps,preprintnumbers,amsmath,amssymb,superscriptaddress,nofootinbib]{revtex4}

\usepackage{graphicx,subfigure,multirow}
 \usepackage{longtable}
\usepackage{dcolumn}
\usepackage{bm}
\usepackage[all]{xy}
\usepackage{color}
\usepackage[usenames,dvipsnames]{xcolor}
\usepackage[unicode=true,pdfusetitle,
 bookmarks=true,bookmarksnumbered=false,bookmarksopen=false,citecolor=Turquoise,
 breaklinks=false,pdfborder={0 0 1},backref=false,colorlinks=true,pdfpagemode=FullScreen]
 {hyperref}

 \newcommand{\lsim}{{\;\raise0.3ex\hbox{$<$\kern-0.75em\raise-1.1ex\hbox{$\sim$}}\;}}
\newcommand{\gsim}{{\;\raise0.3ex\hbox{$>$\kern-0.75em\raise-1.1ex\hbox{$\sim$}}\;}}
\newcommand{\beq}{\begin{equation}}
\newcommand{\eeq}{\end{equation}}
\newcommand{\bea}{\begin{eqnarray}}
\newcommand{\eea}{\end{eqnarray}}
\mathchardef\minus="002D

\def\beq{\begin{equation}}
\def\eeq{\end{equation}}
\def\bea{\begin{eqnarray}}
\def\eea{\end{eqnarray}}
\def\bit{\begin{itemize}}
\def\eit{\end{itemize}}

\def\baa{\begin{array}}
\def\eaa{\end{array}}

\def\misse{E\hspace{-0.25cm}/_{T}}

\begin{document}

\title{\boldmath 
Kinematic Discrimination of $tW$ and $t\bar{t}$ Productions Using Initial State Radiation
}
\author{Doojin Kim}
\affiliation{Department of Physics, University of Florida, Gainesville, FL 32611, USA}
  
\author{Kyoungchul Kong}
\affiliation{Department of Physics and Astronomy, University of Kansas, Lawrence, KS 66045, USA}

\begin{abstract}
Production of a single top quark provides excellent opportunity for understanding top quark physics and Cabibbo-Kobayashi-Maskawa structure of the quark sector in the Standard Model.
Although an associated production with a $b$-quark has already been observed at the Tevatron in 2009, 
a single top production in association with a $W$ gauge boson has not been observed till 2014 at the LHC, 
where pair production of the top quark serves as the dominant background.
Due to the kinematic similarity between $tW$ and the dominant background, it is challenging to find suitable kinematic variables that offer good signal-background separation, 
which naturally leads to the use of multivariate methods.
In this paper, we investigate kinematic structure of $tW+j$ channel using $M_{T2}$ and invariant mass variables, and find 
that $tW +j$ production could well be separated from $t\bar t$ production with high purity at a low cost of statistics when utilizing these kinematic correlations.

\bigskip

\end{abstract}


\maketitle


\section{\label{sec:intro} Introduction} 
The research program at the Large Hadron Collider (LHC) has been greatly successful in the sense that it {\it not} only discovered a new scalar state~\cite{Aad:2012tfa,Chatrchyan:2012ufa}, which is consistent with the Higgs boson in the Standard Model (SM), but {\it re}discovered the SM with great precision. Among the precision studies, the top quark ($t$) has received a particular attention as it is, and also as a window to new physics discovery. In fact, the LHC, dubbed as ``top factory'', is capable of copiously producing top quarks in pair via the strong interaction. Although mediated by the electroweak interaction, the production rate of a single top quark is quite sizable due to a large center of mass energy so that the LHC can provide with an ideal environment to study the single top modes as well. In the SM, the relevant production cross section of a single top is directly proportional to squaring one of the Cabibbo-Kobayashi-Maskawa (CKM) matrix elements, $V_{tb}$, so that single top channels serve as a way to measure the parameter. On top of this parameter measurement, their cross section measurement is also sensitive to various new phenomena such as forth-generation models and models with flavor-changing neutral currents~\cite{Tait:2000sh}. 

The production of a single top through $s$-channel and $t$-channel $W$ gauge boson exchanges had been observed, at the 5.0 standard deviation level of significance, separately by D0~\cite{Abazov:2009ii} and by CDF~\cite{Aaltonen:2009jj}, whereas the associated production of a single top with a $W$ gauge boson (henceforth denoted by $tW$) had too small a cross section to be observed at the Tevatron. Nevertheless, the discovery of the $tW$ channel becomes of great importance in the sense of 1) a way of confirming the SM in the top sector, 2) a way or cross-check of $|V_{tb}|$ measurement, and 3) a possible link to new physics searches such as bottom partners \cite{Aad:2013rna,Nutter:2012an}. The LHC experiment has been able to reach a sufficient production cross section to see the $tW$ mode~\cite{Chatrchyan:2014tua,ATLAS} only in five years after the discovery of $s$-channel and $t$-channel single top modes, and 
the combination of their cross section measurements can be found in Ref. \cite{CMS:2014efa}.
The ATLAS and CMS collaborations have devoted a lot of effort to develop a variety of sophisticated multivariate techniques that take advantage of the differences in the kinematic distributions between the relevant signal and backgrounds, i.e., the method of Boost Decision Tree (BDT) for the CMS and the method of Multi-Variate Analysis (MVA) for the ATLAS. Yet, there is {\it no} single kinematic variable that serves the reasonable separation between the signal and backgrounds. 

\begin{figure}[t]
\centering
\includegraphics[scale=0.31]{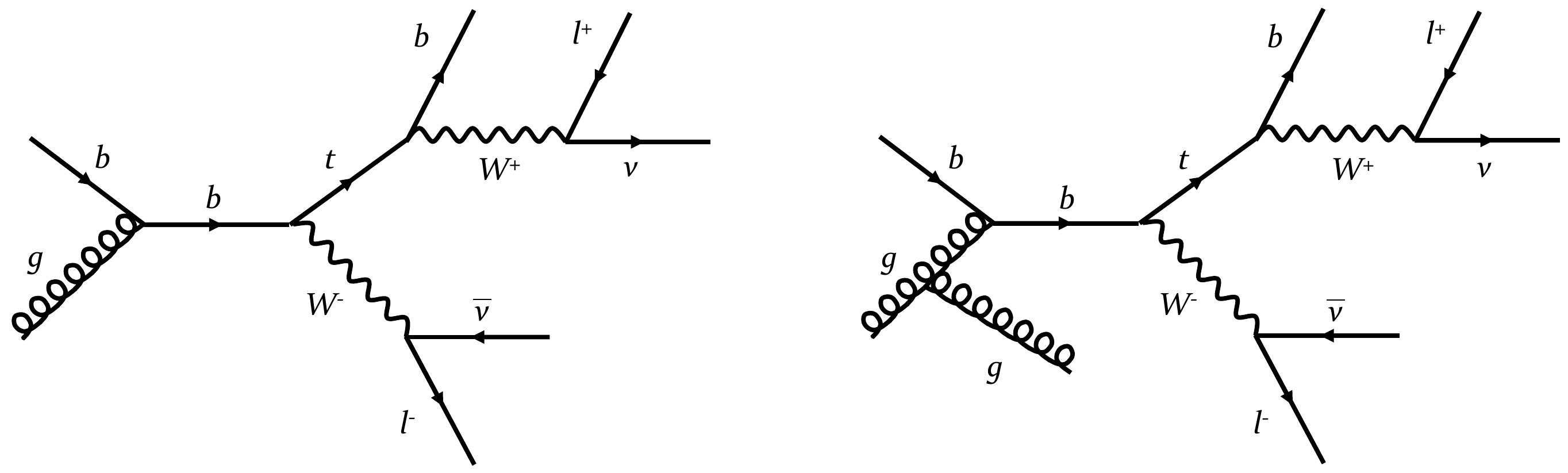}
\caption{\label{fig:topology} A sample Feynman diagram of the associated production of a single top with a $W$ gauge boson and their subsequent decay (left panel) and one with an extra jet attached (right panel). }
\end{figure}

The signal channel is defined by the process shown in the left panel of Figure~\ref{fig:topology}, while the major background to this channel is identified as the ordinary pair-produced top quarks for which one of the bottom quarks is missed (typically by transverse momentum and pseudo-rapidity acceptance). 
Although the corresponding probability may not be large, 
the overwhelming production rate of $t\bar{t}$ can give rise to a sizable background to the signal process.
This expectation is clearly reflected in the CMS analysis of Ref.~\cite{CMS:2014efa}. 
Their signal region is defined by exactly one $b$-tagged jet (together with two $W$'s). 
Although the signal region predominantly contains $tW$ and $t\bar{t}$ events (after their selection criteria), $t\bar{t}$ is still $\sim 5$ times larger than $tW$, (again motivating the adoption of multivariate techniques as a posterior data analysis scheme).


It is interesting to compare the kinematic feature between the $tW$ and the $t\bar{t}$ systems. First of all, the bottom quark comes from the decay of a top quark with a $W$ gauge boson for both signal and background processes, i.e., the typical hardness and the directional preference of the bottom quark are similar. Therefore, its kinematic ensemble for both $tW$ and $t\bar{t}$, e.g., the distribution in the transverse momentum, tends to be close to each other. An analogous argument is readily applicable to the lepton. For both signal and background, it is emitted from a $W$ boson along with a neutrino, so that the typical hardness and the directional preference are anticipated to be similar. Along the line of this observation, it is not surprising that other variables induced from the momenta of $b$-quarks and leptons do not show a reasonable performance in separating the signal and background events. In other words, it is rather difficult to find suited kinematic variables that offer good signal-background discrimination. 

Provided with such a challenging situation, we here propose an alternative kinematic variable-based strategy which could have expedited the observation of the single top mode associated with a $W$ gauge boson. The main idea behind our proposal can be summarized as follows. We basically require an additional jet on top of a bottom-tagged jet, two opposite-signed leptons, and a (large) missing transverse energy in the final state. Such an extra jet can be either $b$-tagged or not, i.e., $2b+\ell^+\ell^-+\misse$ or $1b+1j+\ell^+\ell^-+\misse$, correspondingly. For the latter signal region, we proceed exactly the same analysis as the former, i.e., we treat the additional non-$b$-tagged jet as if it were a bottom-initiated jet. With this requirement, the background restores the regular dileptonic $t\bar{t}$ event topology.\footnote{Of course, one of the two $b$-jets can be either $b$-tagged or not as well.} On the other hand, the signal process comes with a single $b$-quark at the leading order, so that higher order contributions are essential to meet the requirement, i.e., demanding an extra jet to attach to the leading order process. An example diagram is illustrated in the right panel of Figure~\ref{fig:topology}. Unlike the background, the $tW$ with an additional jet has an ill-defined event topology because such a jet is typically from either initial state radiation (ISR) or final state radiation (FSR). We then apply the well-known $M_{T2}$ variable~\cite{Lester:1999tx,Cho:2007qv,Burns:2008va,Kim:2009si} and the conventional invariant mass variable formed by a bottom quark and a lepton, $m_{b\ell}$. While the background yields upper-bounded distributions in those variables, the signal distributions are expected to stretch further beyond the kinematic endpoints of the background, for which the details are dictated by the hardness of the extra jet. It is therefore expected that a large fraction of signal events survive even with kinematic cuts in the $M_{T2}$ and $m_{b\ell}$ 
while the background events are significantly suppressed. 
A related approach has been examined in Ref. \cite{Alwall:2009zu} to solve combinatorial issues with ISR in new physics signals involving jets.
Our approach is different, and with our findings we suggest to use ISR to suppress backgrounds in the given final state for an expedite discovery and precision measurement.

The rest of this paper is organized as follows. In the next section, we briefly review the $M_{T2}$ variable, taking the dileptonic $t\bar{t}$ as a concrete example. In Sec.~\ref{sec:LO}, we discuss behaviors of $t\bar{t}$ and $tW$ in the $M_{T2}$ and $m_{b\ell}$ variables with the requirement of $1b+2\ell+\misse$. 
We then re-examine their behaviors in those variables with an additional jet requirement in Sec.~\ref{sec:NLO}. Sec.~\ref{sec:discussion} is reserved for our discussions and outlook. 

\section{\label{sec:review} A review on the $M_{T2}$ variable}	

\begin{figure}[t]
\centering
\includegraphics[scale=0.53]{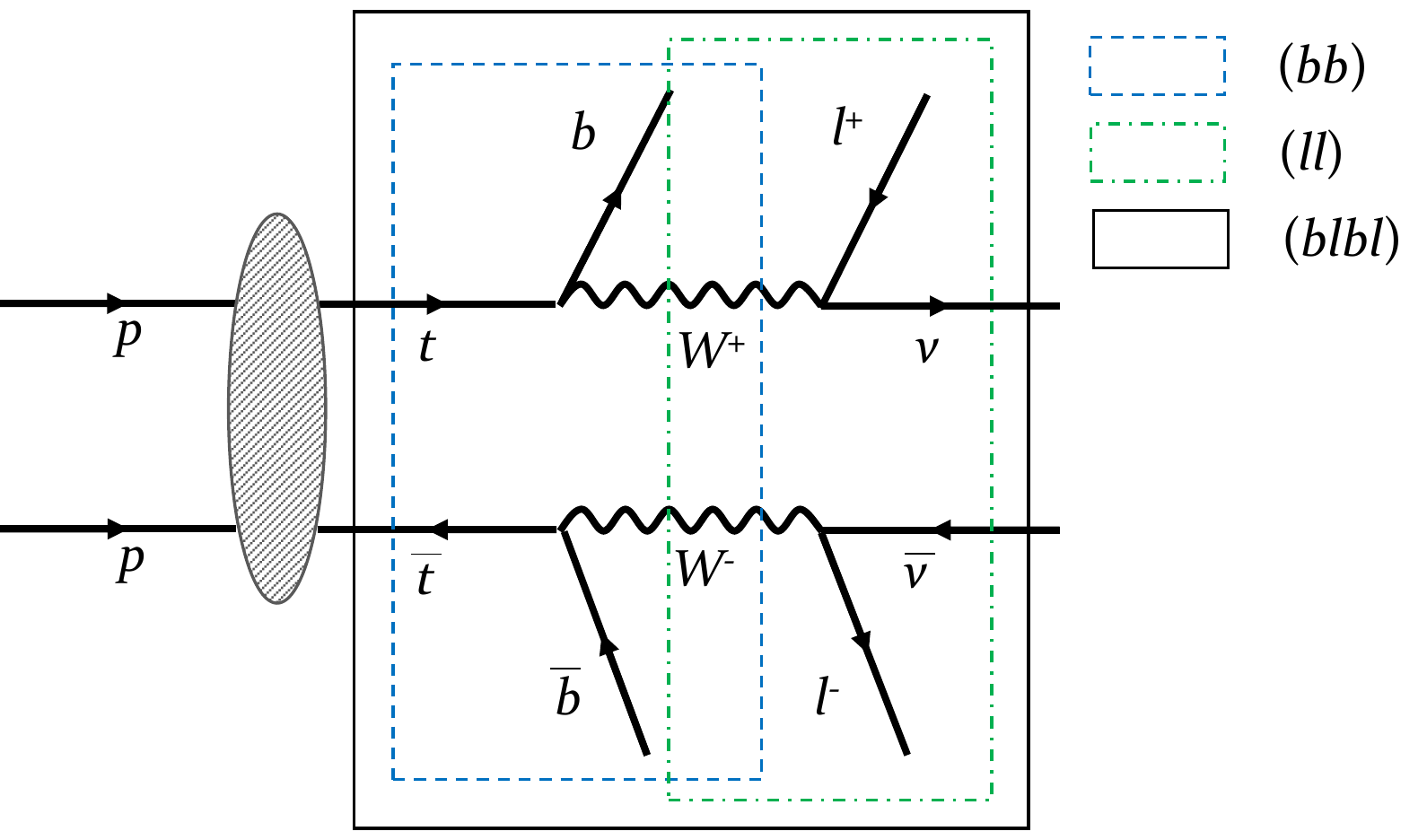}
\caption{\label{fig:ttbartopo} The dileptonic $t\bar{t}$ decay process with the corresponding symmetric subsystems explicitly specified. The blue dotted, green dot-dashed, and black solid boxes indicate subsystems $(bb)$, $(\ell\ell)$, and $(b\ell b\ell)$, respectively.}
\end{figure}

$M_{T2}$ and $m_{b\ell}$ variables are well-motivated especially 
for a cascade decay of a heavy particle including two-step two-body decays such as the top decay, and therefore, it makes sense to investigate them for the $tW$ case. While $m_{b\ell}$ is (relatively) well-known, the $M_{T2}$ variable has non-trivial and less familiar features. 
In this sense, we provide a brief review on the $M_{T2}$ variable that is employed for the analyses in the following sections. For concreteness of the discussion later, we take the event topology defined by the pair-produced top quarks which subsequently decay dileptonically (see also Figure~\ref{fig:ttbartopo}):
\bea
t\bar{t} \rightarrow bW^+ \bar{b}W^- \rightarrow b\ell^+\nu \bar{b}\ell^-\bar{\nu} \, .
\eea
We also take the decay sequence initiated by the top quark as the first decay side, while that by the anti-top quark as the second decay side solely for convenience. 

The $M_{T2}$ variable was originally proposed as a simple generalization of the well-known transverse mass to the case where each of the pair-produced heavier particles decays into an invisible particle along with a visible state~\cite{Lester:1999tx,Cho:2007qv,Burns:2008va,Kim:2009si}. Since the total missing transverse momentum $\mpt$ is shared by the two invisible particles, its formal definition is given by a minimization of the maximum of the two transverse masses ($M_T^{(1)}$ and $M_T^{(2)}$) in each decay chain over the transverse components of the invisible momenta (denoted by $\vec{q}_{T}^{(1)}$ and $\vec{q}_{T}^{(2)}$), subject to the $\mpt$ constraint, i.e., the total sum of the transverse momenta should identically vanish:
\bea
M_{T2}(\tilde{m})&\equiv & \min_{\vec{q}_{T}^{\,(1)},\vec{q}_{T}^{\,(2)}}\left\{ \max\left[ M_T^{(1)}(\vec{q}_{T}^{\,(1)},\tilde{m}),\;M_T^{(2)}(\vec{q}_{T}^{\,(2)},\tilde{m}) \right]\right\} \, , \nonumber\\
0 &=&\vec{q}_{T}^{\,(1)}+\vec{q}_{T}^{\,(2)}-\mpt \, ,  \label{eq:MT2def}
\eea
where $\tilde{m}$ denotes the hypothetical/test mass parameter for the invisible particles and the superscripted numbers indicate the associated decay side. When more than one visible particle is involved in each decay chain, then one can define $M_{T2}$ in various subsystems~\cite{Burns:2008va} which can be further categorized into symmetric and asymmetric subsystems whether or not both $M_T^{(i)}$'s $(i=1,2)$ are constructed in the same fashion. For the case of the $t\bar{t}$ system, there are three {\it symmetric} subsystems which are henceforth denoted by $(bb)$, $(\ell\ell)$, and $(b\ell b\ell)$ subsystems as per the visible particles associated with the subsystem under consideration. We explicitly delineate those three subsystems in Figure~\ref{fig:ttbartopo}, and the operational difference among them is summarized below:
\begin{itemize}
\item For the $(b\ell b\ell)$ subsystem, the transverse masses for the top quarks are minimized with the neutrinos considered as invisible particles.
\item For the $(\ell\ell)$ subsystem, the transverse masses for the $W^{\pm}$ are minimized with the neutrinos considered as invisible particles. The visible momenta for the bottom quarks are considered as upstream momenta.
\item For the $(bb)$ subsystem, the transverse masses for the top quarks are minimized with the $W^{\pm}$ considered as invisible particles. The visible momenta for the leptons are considered as downstream momenta so that they are treated invisibly. 
\end{itemize}   
Since the neutrino plays a role of the invisible particle in the $(b\ell b\ell)$ and $(\ell\ell)$ subsystems, the relevant test mass is typically assumed to be 0 GeV as per the SM neutrino mass. Analogously, for the $(bb)$ subsystem, the relevant test mass is typically assumed to be 80 GeV as per the mass of the $W$ gauge boson. 

\begin{figure}[t]
\centering
\includegraphics[scale=0.53]{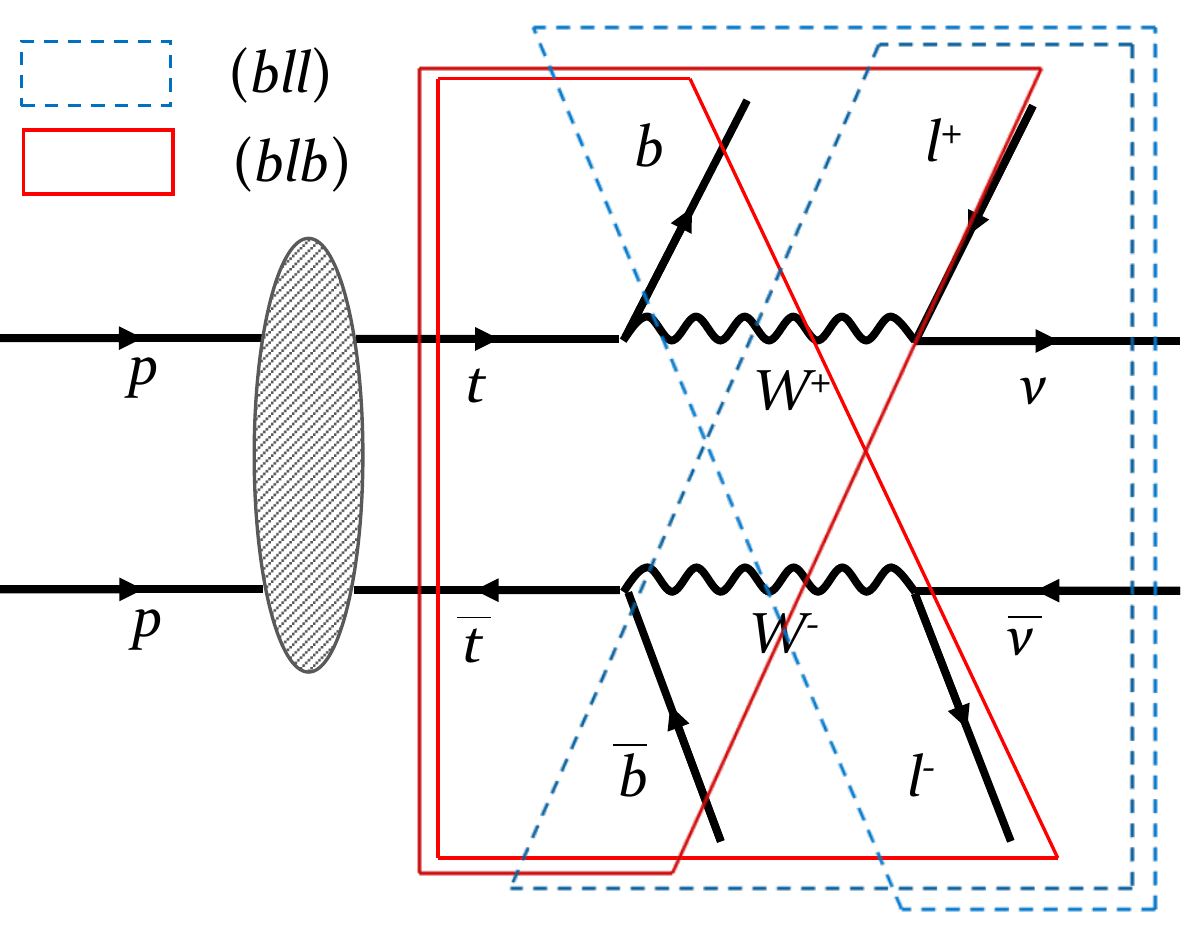}  \\
\includegraphics[scale=0.53]{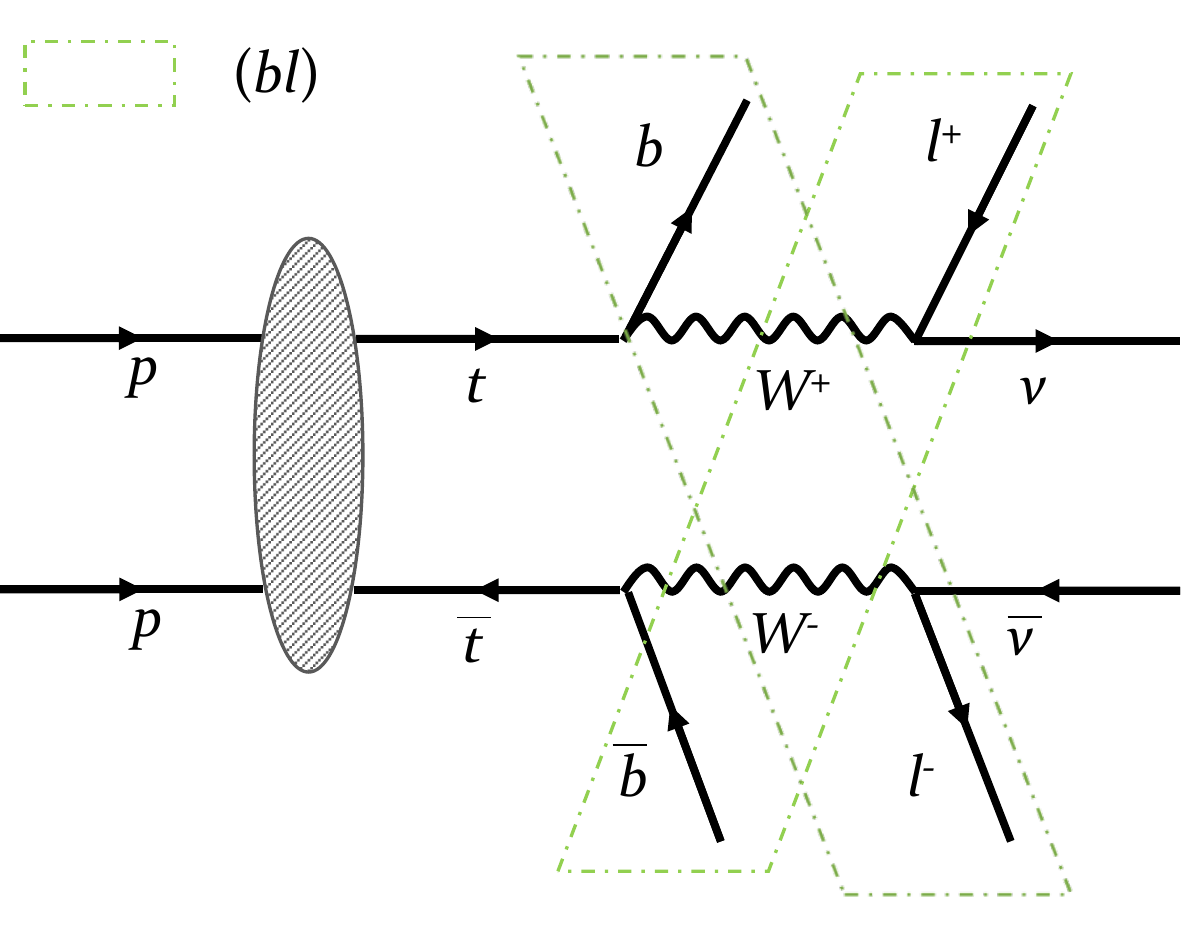}
\caption{\label{fig:ttbartopoasym} The dileptonic $t\bar{t}$ decay process with the corresponding asymmetric subsystems explicitly specified. The blue dotted and red solid boxes in the left panel and the green dot-dashed box in the right panel indicate subsystems $(b\ell\ell)$, $(b\ell b)$, and $(b\ell)$, respectively. }
\end{figure}
Similar constructions can be performed for the {\it asymmetric} subsystems \cite{Konar:2009qr}. In this case, there arise three different subsystems denoted by $(b\ell\ell)$, $(b\ell b)$, and $(bl)$ again named after the visible particles associated with the subsystem of interest. The corresponding subsystems are explicitly delineated in Figure~\ref{fig:ttbartopoasym}, and the operational difference among them is explained below:
\begin{itemize}
\item For the $(b\ell\ell)$ subsystem, the transverse masses for the top quark in one decay side and the $W^{\pm}$ in the other decay side are minimized with the neutrinos considered as invisible particles. The visible momentum for the remaining bottom quark is considered as upstream momenta.
\item For the $(b\ell b)$ subsystem, the transverse masses for the top quarks are minimized with the neutrino in one decay side and the $W^{\pm}$ in the other decay side considered as invisible particles. The visible momentum for the remaining lepton is considered as downstream momenta so that it is treated invisibly.
\item For the $(b\ell)$ subsystem, the transverse masses for the top quark in one decay side and the $W^{\pm}$ in the other decay side are minimized with the neutrino in one decay side and the $W^{\pm}$ in the other decay side considered as invisible particles. The visible momenta for the remaining bottom quark and lepton are considered as upstream and downstream momenta, respectively, the latter of which is treated invisibly.
\end{itemize}
Since the neutrino is considered as the invisible particle in both decay sides for the $(b\ell\ell)$ subsystem, the relevant test mass is typically assumed to be 0 GeV as per the SM neutrino mass. On the contrary, in the other two subsystems, two different particle species take over the role of invisible particles so that two different test masses can be imposed, accordingly, i.e., 0 GeV and 80 GeV as per the masses of the SM neutrino and $W$ gauge boson, depending on the subsystem of interest.

One noteworthy fact is that the associated $M_{T2}$ distributions are bounded above by the mass of the decaying particle.\footnote{Strictly speaking, this statement is true only if the actual event comes from a well-defined decay topology. We will see that this is {\it not} the case for our signal process, i.e., $tW+j$ from an {\it ill}-defined decay topology.} In fact, the analytic expressions for the kinematic endpoints can be written in terms of the mass parameters involved in the decay process~\cite{Lester:1999tx,Cho:2007qv,Burns:2008va}, and interestingly enough, if the test masses are the same as the masses of invisible particles in the relevant subsystem, the maximum $M_{T2}$ value is the same as the heavier of the actual masses of the particles whose transverse masses are minimized. For our $t\bar{t}$ example, subsystems $(b\ell b\ell)$, $(bb)$, $(b\ell\ell)$, $(b\ell b)$, and $(b\ell)$ simply return the top quark mass while subsystem $(\ell\ell)$ simply returns the $W$ mass if each of the test masses is imposed correspondingly. 

\section{\label{sec:LO} $tW$ at the leading order: existing analyses}

We first discuss collider signatures of dileptonic $tW$ channel at the leading order together with a brief review on the corresponding experimental measurements conducted by CMS/ATLAS collaborations~\cite{Chatrchyan:2014tua, ATLAS}. For more concrete discussions later on, Monte Carlo event samples of $t\bar{t}$ and $tW$ including realistic effects such as detector resolutions have been prepared. For both signal ($t W$) and background ($t\bar t$), the parton level events at the leading order are generated by \texttt{MadGraph\_aMC@NLO}~\cite{Alwall:2014hca} in conjunction with parton distribution functions given by \texttt{NNPDF23}~\cite{Ball:2012cx} that is the default of \texttt{MadGraph\_aMC@NLO}. 
Both top quark and $W$ boson are forced to decay inside \texttt{MadGraph\_aMC@NLO} to include the spin-correlation and off-shell effects.
The outcomes ($t\bar t$ and $t W$ events) from the parton event generator are subsequently fed to \texttt{Pythia6.4}~\cite{Sjostrand:2006za} for the showering and hadronization.
Then those events are further processed to \texttt{Delphes3}~\cite{deFavereau:2013fsa} for describing the detector effects. All the simulation is done with a proton-proton collider of $\sqrt{s}=8$ TeV and an input top mass of 173 GeV. 
Note that here we do not simulate signal and background processes with extra radiation (e.g., $t\bar{t}+j$) at the generation level, as the showering by \texttt{Pythia} module can effectively take care of the relevant diagrams~\cite{Corke:2010zj,Corke:2010yf,Mrenna}.

\begin{figure*}[th]
\centerline{
\includegraphics[scale=0.6]{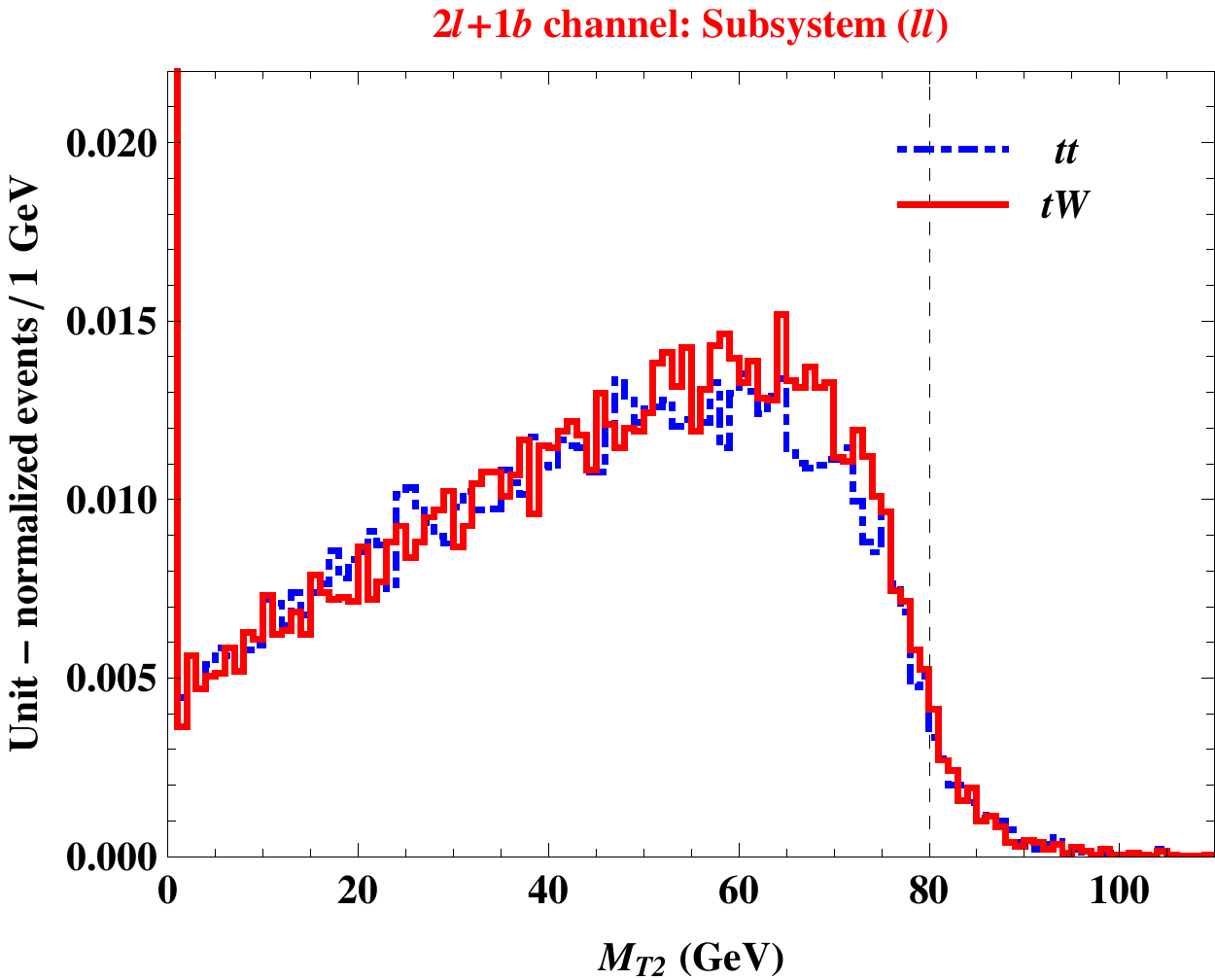}
\hspace{0.8cm}
\includegraphics[scale=0.6]{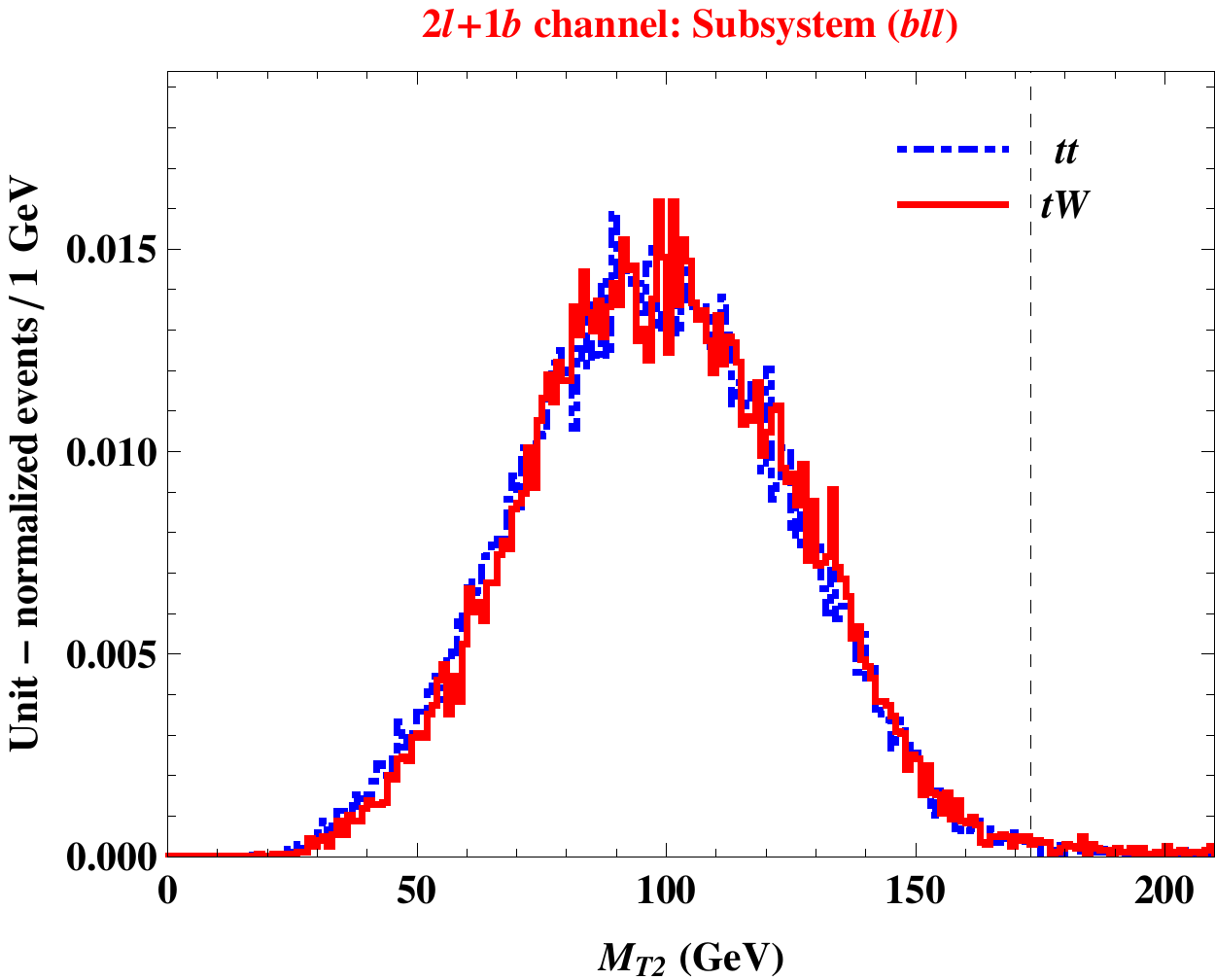} }
\vspace*{0.2cm}
\centerline{
\includegraphics[scale=0.6]{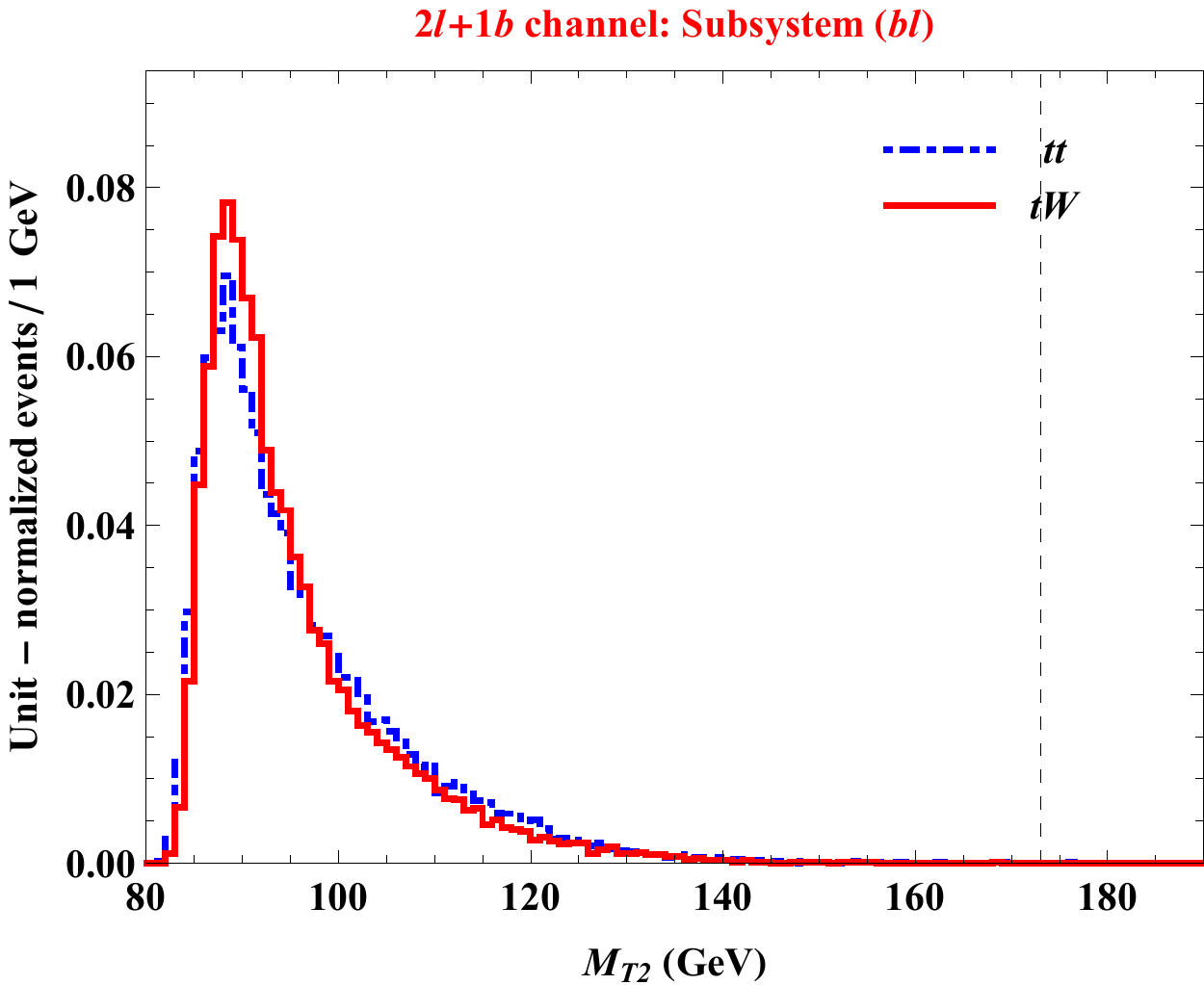}
\hspace{0.8cm}
\includegraphics[scale=0.6]{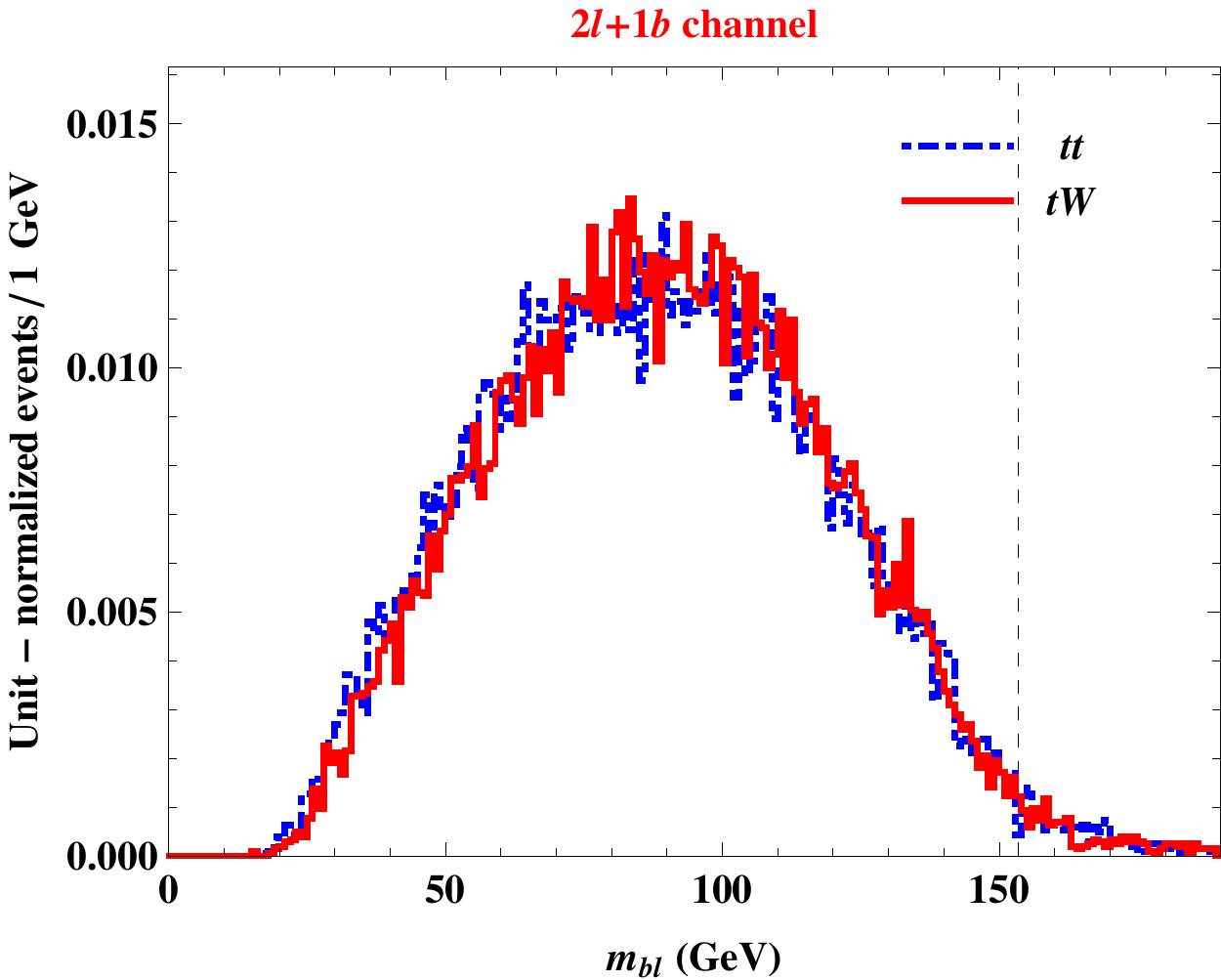}}
\caption{\label{fig:Inv2l1b} $M_{T2}$ distributions in various subsystems -- subsystem $(\ell\ell)$ (upper-left panel), subsystem $(b\ell\ell)$ (upper-right panel), and subsystem $(b\ell)$ (lower-left panel) -- and the $m_{b\ell}$ distributions (lower-right panel) for $t\bar{t}$ and $tW$ events. The distributions are plotted with the events passing the selection criteria listed in Eqs.~(\ref{eq:cutfirst}) through~(\ref{eq:cutlast}). The combinatorics arising in $M_{T2}$ for subsystems $(b\ell\ell)$ and $(b\ell)$ and $m_{b\ell}$ is treated by choosing the smaller of the two possible values in each variable. The test mass for $M_{T2}$ is 0 GeV for subsystems $(\ell\ell)$ and $(b\ell\ell)$, while for subsystem $(b\ell)$ 0 GeV and 80 GeV are imposed for the lepton side and the bottom side, respectively. The dashed lines indicate the expected endpoints of the $t\bar{t}$ system. }
\end{figure*}

Given the final state defined by the dileptonic $tW$ at the leading order, i.e., $b\ell^{+}\ell^-+\misse$ with $\ell$ being either $e$ or $\mu$, several SM processes can give rise to the same visible final state. It turns out that among them dileptonic $t\bar{t}$ is the dominant background where one of $b$-quarks is lost, and therefore, we focus on the comparison between the two processes throughout this paper. To be mostly left with $tW$ and $t\bar{t}$ events, we closely follow the event selection scheme employed in Ref.~\cite{Chatrchyan:2014tua}, among which the key criteria are enumerated below: 
\begin{eqnarray}
&&\hspace{-1cm}\bullet \; N_{\ell}=2 \hbox{ with opposite electric charges, } \nonumber \\
&&\hspace{-1cm}p_T^{e,\;\mu}>10 \hbox{ GeV and } |\eta^{e(\mu)}|<2.5 ~(2.4), \label{eq:cutfirst}\\
&&\hspace{-1cm}\bullet \; \misse > 50 \hbox{ GeV for the same flavor channels}, \\
&&\hspace{-1cm}\bullet \; m_{\ell\ell}>20 \hbox{ GeV and } |m_{\ell\ell}-m_Z|>10 \hbox{ GeV} , \label{eq:cutmiddle} \\
&&\hspace{-1cm}\bullet \; N_j=0 \hbox{ while } N_b=1,\; p_T^{j(b)}>20 ~(30) \hbox{ GeV and } \nonumber \\
&&\hspace{-1cm} |\eta^{j(b)}|<4.9~(2.4), \label{eq:cutlast}  
\end{eqnarray}
where $N_{\ell}$ and $N_{j(b)}$ denote the number of selected leptons and jets ($b$-tagged jets), respectively, and $\misse$ is defined as $|\sum_i \vec{p}_T^{\,i} |=|-\mpt|$ with $i$ being all detected particle species. Jets are formed by the anti-$k_t$ algorithm~\cite{Cacciari:2008gp} together with a radius parameter $R=0.5$, and the $b$-tagging efficiency is hardwired to be 70~\%, while the light quark jets are mis-tagged by 1\% \cite{Chatrchyan:2014tua}
\footnote{In Ref.~\cite{Chatrchyan:2012jua}, the CMS collaboration observed similar tagging efficiency and mis-tag rate in events from multijet and top-quark pair productions.}.  
A jet is tagged as a $b$-jet if its direction lies in the acceptance of the tracker and if it is associated to a parent $b$-quark~\cite{deFavereau:2013fsa}.

Having the events passing the above-given selection cuts, we first show that conventional kinematic variables such as $M_{T2}$ for three available subsystems and $m_{b\ell}$ would not help us separate the $tW$ events from the $t\bar{t}$ ones. The relevant distributions are exhibited in the upper-left panel ($M_{T2}$ in the $(\ell\ell)$ subsystem), the upper-right panel ($M_{T2}$ in the $(b\ell\ell)$ subsystem), the lower-left panel ($M_{T2}$ in the $(b\ell)$ subsystem), and the right panel ($m_{b\ell}$) of Figure~\ref{fig:Inv2l1b}. Speaking of the $M_{T2}$ variables in various subsystems, we see that both of $t\bar{t}$ (blue dashed histograms) and $tW$ (red solid histograms) develop similar distributions in them. For the case of $t\bar{t}$, the distribution in each subsystem is nothing but the one anticipated in the respective subsystem, and therefore, the associated kinematic endpoint is expected to be the same as the $W$ gauge boson mass ($M_{T2}$ of subsystem $(\ell\ell)$) or the top quark mass ($M_{T2}$ of subsystems $(b\ell\ell)$ and $(b\ell)$) with test masses imposed correspondingly as mentioned earlier~\cite{Burns:2008va}. The theoretical endpoints are indicated by black dashed lines, and we see that most of $t\bar{t}$ events are populated below them as expected. The small overflow in the $M_{T2}$ distributions for the $(\ell\ell)$ and $(b\ell\ell)$ subsystems is due to various sources such as mis-measurement of $\misse$ and parton showering/fragmentation (see, for example, Ref.~\cite{Kim:2014ana} for more systematic study on the effect of those sources). On the other hand, for the $M_{T2}$ distribution in the $(b\ell)$ subsystem, it is hard to find out kinematic configurations corresponding to the relevant endpoint so that the distribution does not reach the expected endpoint. When it comes to the signal process, in some sense, the final state of $tW$ does not differ from that of $t\bar{t}$. For example of the $(\ell\ell)$ subsystem, while the $M_{T2}$ for $tW$ can be interpreted as the one applied to the situation where $W$ gauge bosons are pair-produced with a non-zero transverse upstream momentum given by a bottom quark, the net upstream momentum for $t\bar{t}$ is defined by a vector sum of the transverse momenta of two bottom quarks. Therefore, both distributions are expected to be upper-bounded by the same endpoint as well as to develop similar shapes up to the details of upstream momenta. A similar analogy is relevant to $M_{T2}$ for the other two subsystems. In this case, however, the $t\bar{t}$ is interpreted as a single top production associated with a $W$ gauge boson with a {\it missing} $b$-jet absorbed into the upstream momentum. Again, signal and background distributions are expected to be bounded above by the same endpoint, and are inclined to exhibit similar shapes up to the details of upstream momenta. From all these observations, we conclude that $M_{T2}$'s in various subsystems are not good signal-background discriminators.

Finally, taking the $m_{b\ell}$ distribution (the lower-right panel of Figure~\ref{fig:Inv2l1b}) into consideration, we see very similar behaviors for both $tW$ (red solid histogram) and $t\bar{t}$ (blue dashed histogram). Here since there exists a two-fold combinatorial ambiguity \cite{Baringer:2011nh}, we keep only the smaller of the two to ensure the boundedness of the $m_{b\ell}$ distributions. For both of them, the kinematic endpoint is dictated by the correct combination, i.e., the invariant mass formed by $b$ and $\ell$ belonging to the decay cascade initiated by the same top quark, so that the expected maximum $m_{b\ell}$ should be identical, that is,
\bea
m_{b\ell}^{\max}=\sqrt{m_t^2-m_W^2} \label{eq:mblend} \, ,
\eea
where all final state particles, i.e., bottom jet, lepton, and neutrino, are assumed massless. Again, the theoretical endpoint is indicated by a black dashed line, while the actual distributions involve a small overflow 
that is mostly stemming from the events where an ISR jet is mis-tagged as a bottom quark-initiated jet and off-shell effects.\footnote{In principle, the NLO corrections may affect the kinematic distributions including $m_{b\ell}$~\cite{AlcarazMaestre:2012vp}. However, we expect that the associated effect is not significant, for example, based on the comparison of some kinematic distributions of $tW$ at LO and NLO in Ref.~\cite{Frixione:2008yi}.} In addition to the correct combinations, even the ensemble of incorrectly-combined $m_{b\ell}$ is anticipated to be similar to each other because the lepton in the wrong combinatorial side is emitted from the common particle species $W$ for both $tW$ and $t\bar{t}$. Of course, there may be a difference between the $W$ from the decay of a top quark and the $W$ in association with a top quark. Our simulation result shown in the lower-right panel of Figure~\ref{fig:Inv2l1b}, however, suggests that such a difference be insignificant.\footnote{We also produced the invariant mass distributions in the larger of the two combinations, and found that $t\bar{t}$ and $tW$ give rise to almost the same spectra.} All these observations above confirm that $m_{b\ell}$ as well is not an ideal kinematic variable for discriminating signal events from background ones.

The poor efficiency in separating $tW$ and $t\bar{t}$ events by using a few simple kinematic variables can motivate to employ a more sophisticated method. 
As a matter of fact, the ATLAS and CMS collaborations have made use of Boost Decision Tree (BDT)~\cite{cslu:esca98mm} for the purpose of rejecting more background events with more signal events retained. 
The BDT is a type of multivariate analysis (MVA), which is a category of analysis methods that combine multiple input variables into a single discriminant. 
A BDT takes a number of input variables (chosen by the analyst) and trains a certain number of decision trees to separate the signal and background based on Monte Carlo samples for each (for both CMS and ATLAS, it was $tW$ vs. $t \bar t$, and other backgrounds were {\it not}  included). To improve signal acceptance and background rejection with reliable performance, the relevant machine-training is ``boosted'' by giving a special weight to the cases where signal events are eventually identified as background events and vice versa. It has served very well the purpose of signal-background separation in the context of $tW$ discovery. However, it is rather difficult to find variables yielding the best sensitivity so as to discriminate $tW$ from backgrounds event-by-event. In addition, the eventual performance highly depends on the training samples, so that the internal procedure is rather obscure.

\section{\label{sec:NLO} $tW$ with Initial State Radiation: an alternative strategy}

Motivated by the challenging situation in separating the signal events from the background ones using simple kinematic variables, we propose an alternative kinematic variable-based strategy of enhancing the relevant signal-over-background. The basic idea behind it is to consider a higher order contribution, that is, a simple attachment of an extra jet (see the right panel of Figure~\ref{fig:topology} as an example event topology). 
The additional jet can be either {\it mis}-tagged as a bottom-initiated jet or not, and we consider both cases separately later on. Hence, we define a couple of signal regions whose final states are characterized by two opposite-signed leptons, a large missing energy, and two (one) $b$-tagged and zero (one) ordinary jets:
\bea
\hbox{Signal region I (SR-I): }&& p p \rightarrow 2b+\ell^+\ell^-+\misse. \\
\hbox{Signal region II (SR-II): }&& p p \rightarrow 1b+1j+\ell^+\ell^-+\misse.
\eea
We particularly emphasize that the discriminating power of $M_{T2}$ and $m_{b\ell}$ can be dramatically improved for $tW$ with an extra jet. For the background process (i.e., $t\bar{t}$), the requirement of SR-I simply retrieves the entire dileptonic decay topology of top pairs so that the associated decay topology is totally {\it well}-defined. For SR-II, even if an extra jet is not $b$-tagged, we expect that the relevant final state of the background comes mostly from the dileptonic top pairs, i.e., the associated decay topology is as well-defined as that of SR-I. On the contrary, for the signal process, an extra jet is typically emitted as initial or final state radiation, and thus the relevant event topology is {\it ill}-defined. The main idea behind the proposed strategy is actually to tackle such a difference. Basically, the distributions of $t\bar{t}$ events in $M_{T2}$ of the six subsystems and $m_{b\ell}$ are bounded above, and their upper bound (i.e., kinematic endpoint) can be easily calculated like the case considered in the previous section. On the other hand, for $tW$, the extra jet coming from ISR can be arbitrarily hard so that the corresponding endpoints in the $M_{T2}$ and $m_{b\ell}$ distributions are completely dictated by the hardness of such an additional jet.  

\subsection{Signal region I: $p p \rightarrow 2b+\ell^+\ell^-+\misse$ \label{sec:sr1}}

\begin{figure*}[t]
\centerline{
\includegraphics[scale=0.62]{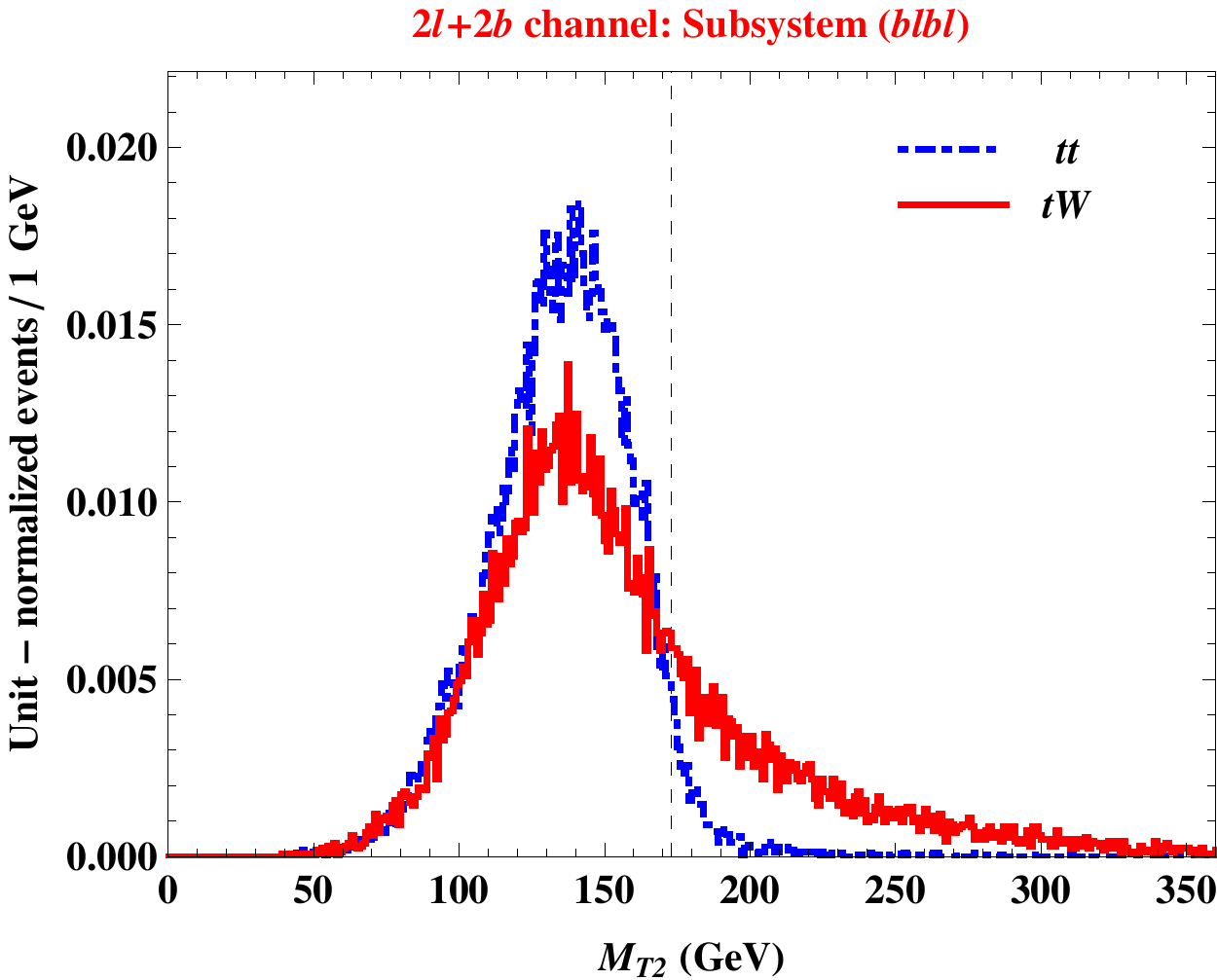}
\hspace{0.5cm}
\includegraphics[scale=0.6]{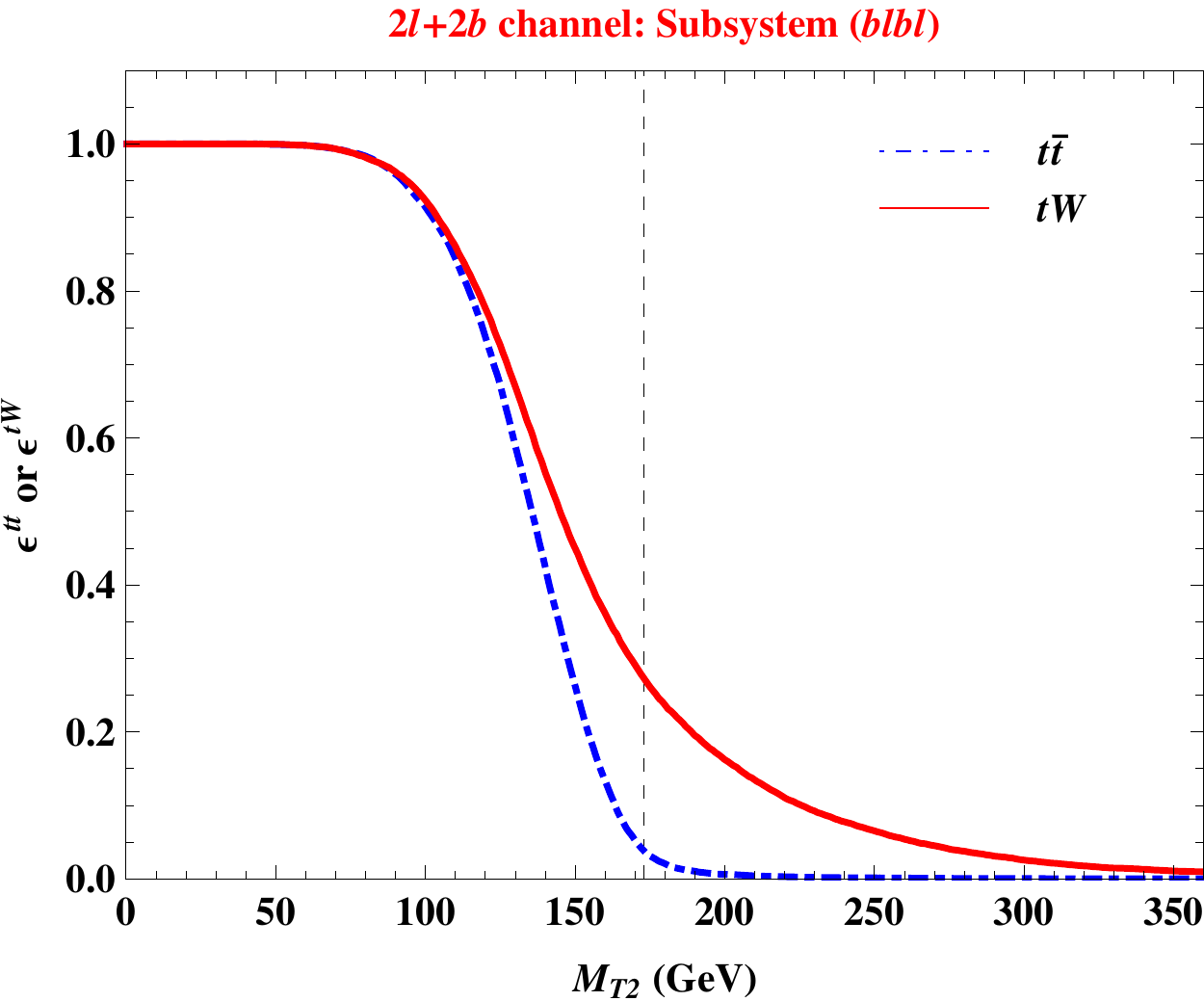}}
\vspace*{0.2cm}
\centerline{
\includegraphics[scale=0.62]{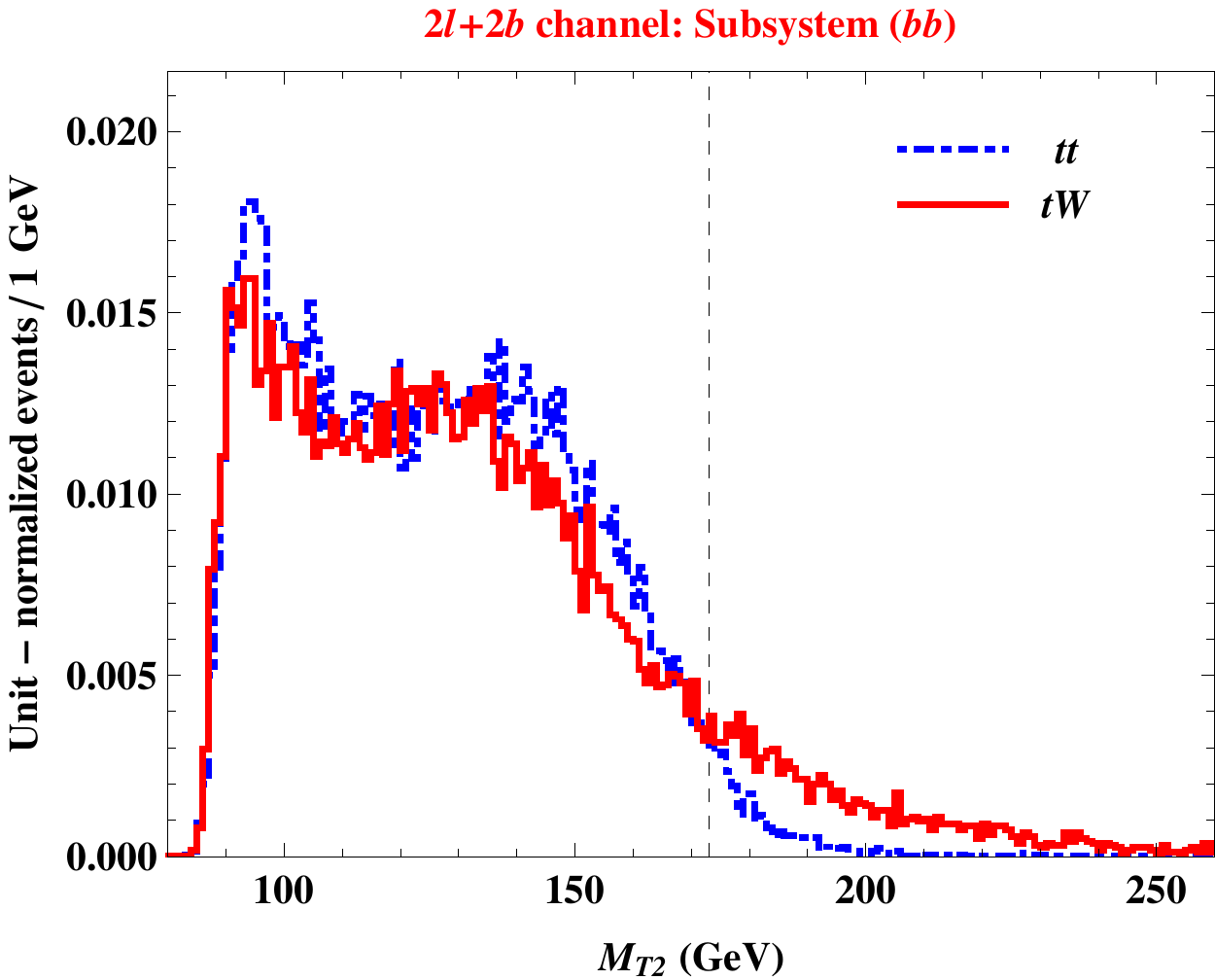}
\hspace{0.5cm}
\includegraphics[scale=0.6]{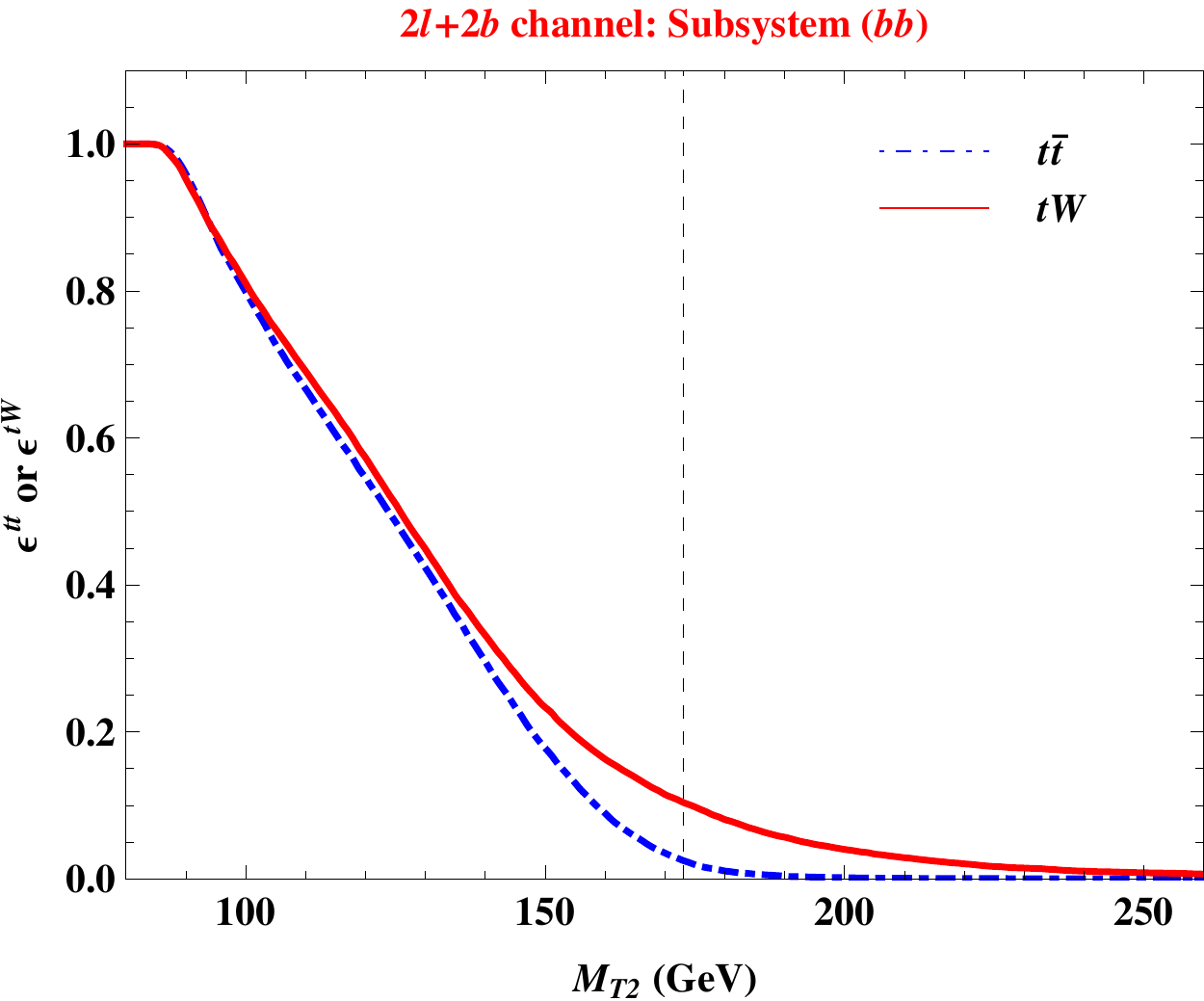} }
\caption{\label{fig:MT2blb}$M_{T2}$ distributions of $t\bar{t}$ and $tW$ events for the $(b\ell b\ell)$ (upper-left panel) and $(bb)$ (lower-left panel) subsystems and
their corresponding selection efficiencies (right panels) in SR-I. The distributions are plotted with the events passing the selection criteria in Eqs.~(\ref{eq:cutfirst}) through~(\ref{eq:cutlast}) with one more $b$-tagged jet is required. The relevant combinatorics arising in the $(b\ell b\ell)$ subsystem is treated by choosing the smaller of the two possible $M_{T2}$ values. The test mass for the ($b\ell b\ell$) subsystem is 0 GeV, while that for the ($bb$) subsystem is 80 GeV. The dashed lines indicate the expected endpoints of the $t\bar{t}$ system. }
\end{figure*}

We begin with the discussion for signal region I, followed by that for signal region II in the next subsection. The event selection scheme for SR-I is the same as Eqs.~(\ref{eq:cutfirst}) through~(\ref{eq:cutlast}) with an additional $b$-tagged jet. Now that we require an additional jet in the final state, the $M_{T2}$ of the $(\ell\ell)$ subsystem is not substantially affected. The extra jet can be absorbed into the upstream momentum with respect to the $(\ell\ell)$ subsystem, i.e., it is simply a redefinition of the ensemble of the upstream momentum that exists in the $(\ell\ell)$ subsystem $M_{T2}$ for the leading order case. On the other hand, the $M_{T2}$ distributions for the other five subsystems show a significant difference between $t\bar{t}$ and $tW$. We first exhibit the unit-normalized $M_{T2}$ distributions of symmetric subsystems in Figure~\ref{fig:MT2blb}: $(b\ell b\ell)$ subsystem in the upper-left panel and $(bb)$ subsystem in the lower-left panel. The blue dashed and the red solid histograms correspond to $t\bar{t}$ and $tW$ systems, respectively. Here the test masses are chosen to be 0 GeV and 80 GeV for the $(b\ell b\ell)$ and the $(bb)$ subsystems, correspondingly, while the black dashed lines denote the the theory predictions for the $M_{T2}$ endpoints of the $t\bar{t}$ system. The well-known two-fold ambiguity arising in the $(b\ell b\ell)$ subsystem is treated by taking the smaller of the two possible $M_{T2}$ values. We clearly see that most of the $t\bar{t}$ events are confined below the expected kinematic endpoint, whereas a large fraction of the $tW$ events exceed the kinematic endpoints for the $t\bar{t}$ system. Therefore, if one sets the cut near the kinematic endpoint, i.e., keeping the event whose $M_{T2}$ value is greater than the cut, one can reject most of the background events with many signal events retained. 

Given the way of keeping or rejecting events with respect to a fixed $M_{T2}$ cut, the associated efficiencies can be defined as a ratio of the number of events passing the cut to the total number of events: 
\begin{equation}
\epsilon^{t\bar t/tW} \equiv \frac{N^{t\bar t/tW}({\rm after}~ M_{T2} ~{\rm cut})}{N^{t\bar t/tW}({\rm before}~ M_{T2} ~{\rm cut})} \, .
\end{equation}
Note that efficiencies with the invariant mass ($m_{b\ell}$) will be defined in a similar fashion.
The right panels of Figure~\ref{fig:MT2blb} demonstrate the associated efficiency curves for the $t\bar{t}$ and $tW$ in the $M_{T2}$ cuts. 
They clearly show that the signal efficiency, $\epsilon^{tW}$ (red solid curves) overwhelms the background efficiency, $\epsilon^{tt}$ (blue dashed curves) as the cuts are close to or beyond the $t\bar{t}$ kinematic endpoints (black dashed lines). 

\begin{figure*}[t]
\centerline{
\includegraphics[scale=0.565]{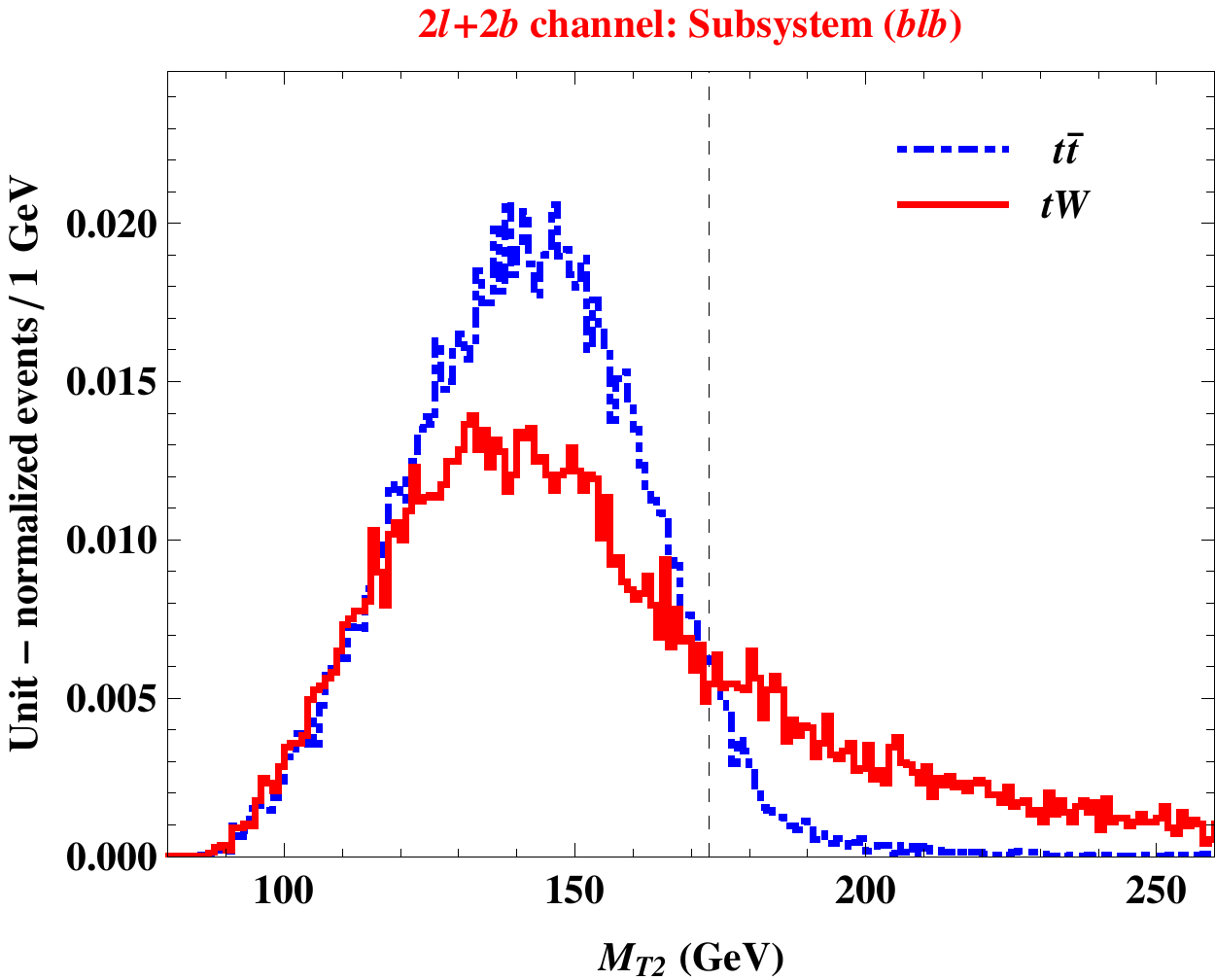}
\hspace{0.4cm}
\includegraphics[scale=0.55]{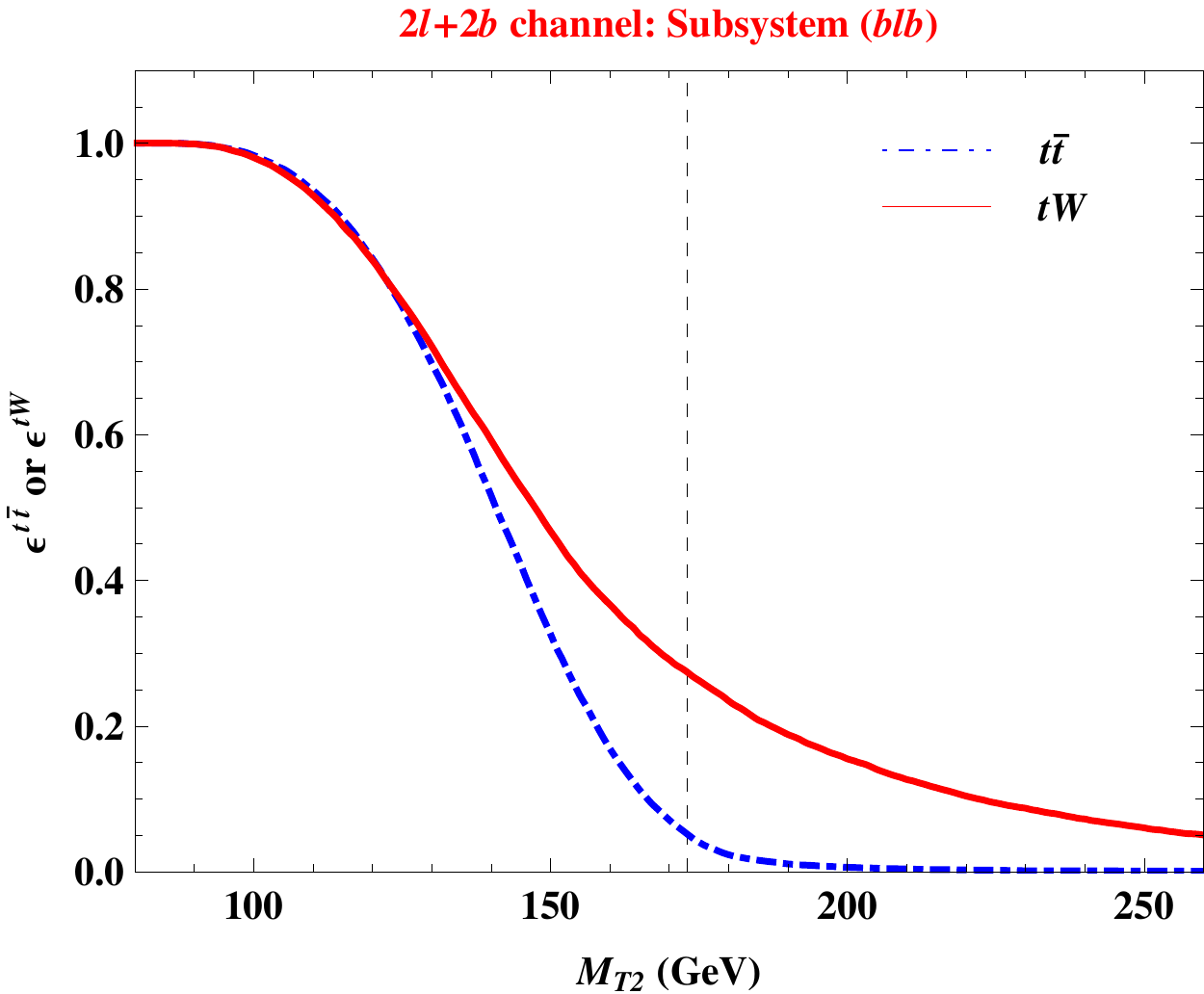}  }
\vspace{0.1cm}
\centerline{
\includegraphics[scale=0.565]{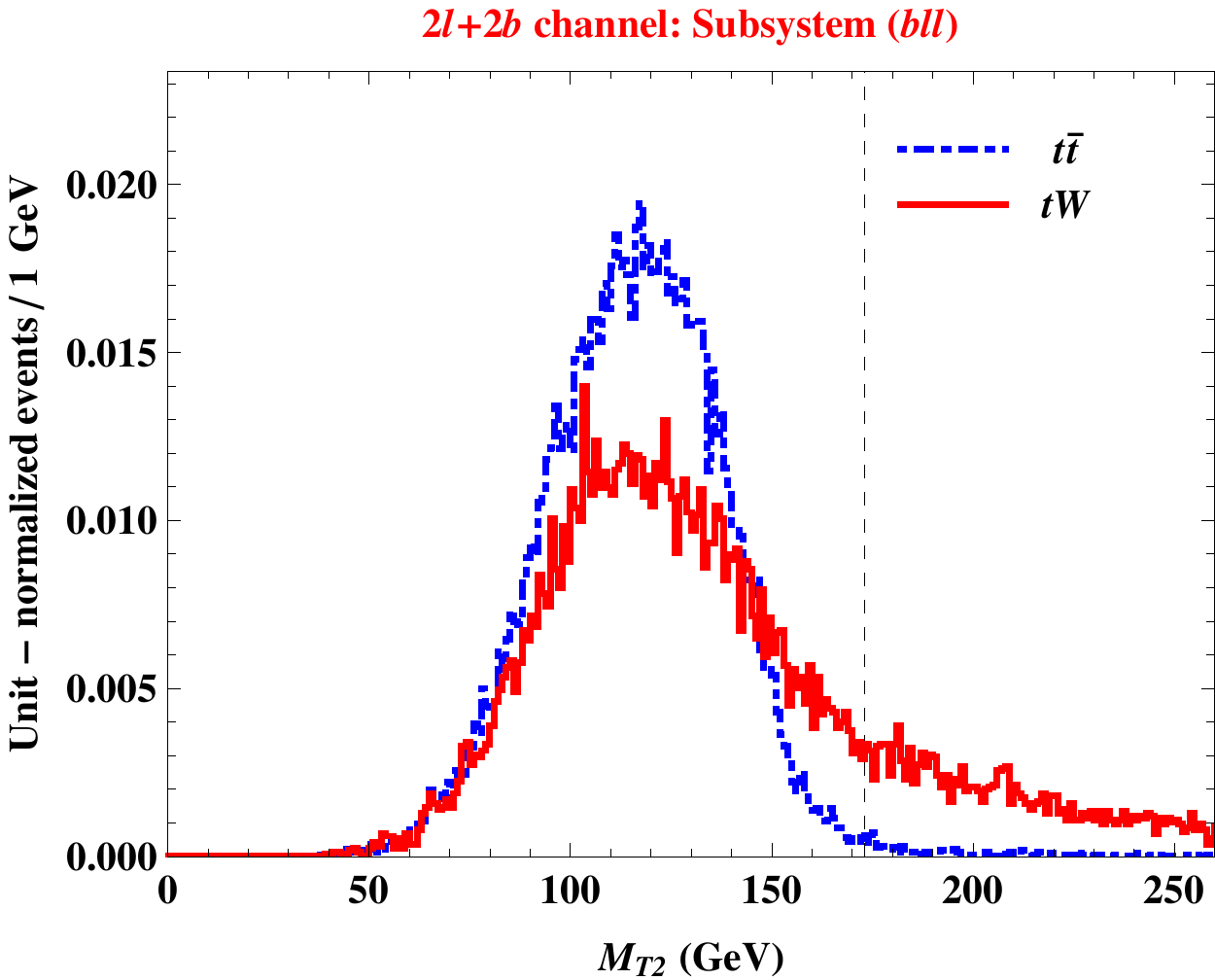}
\hspace{0.4cm}
\includegraphics[scale=0.55]{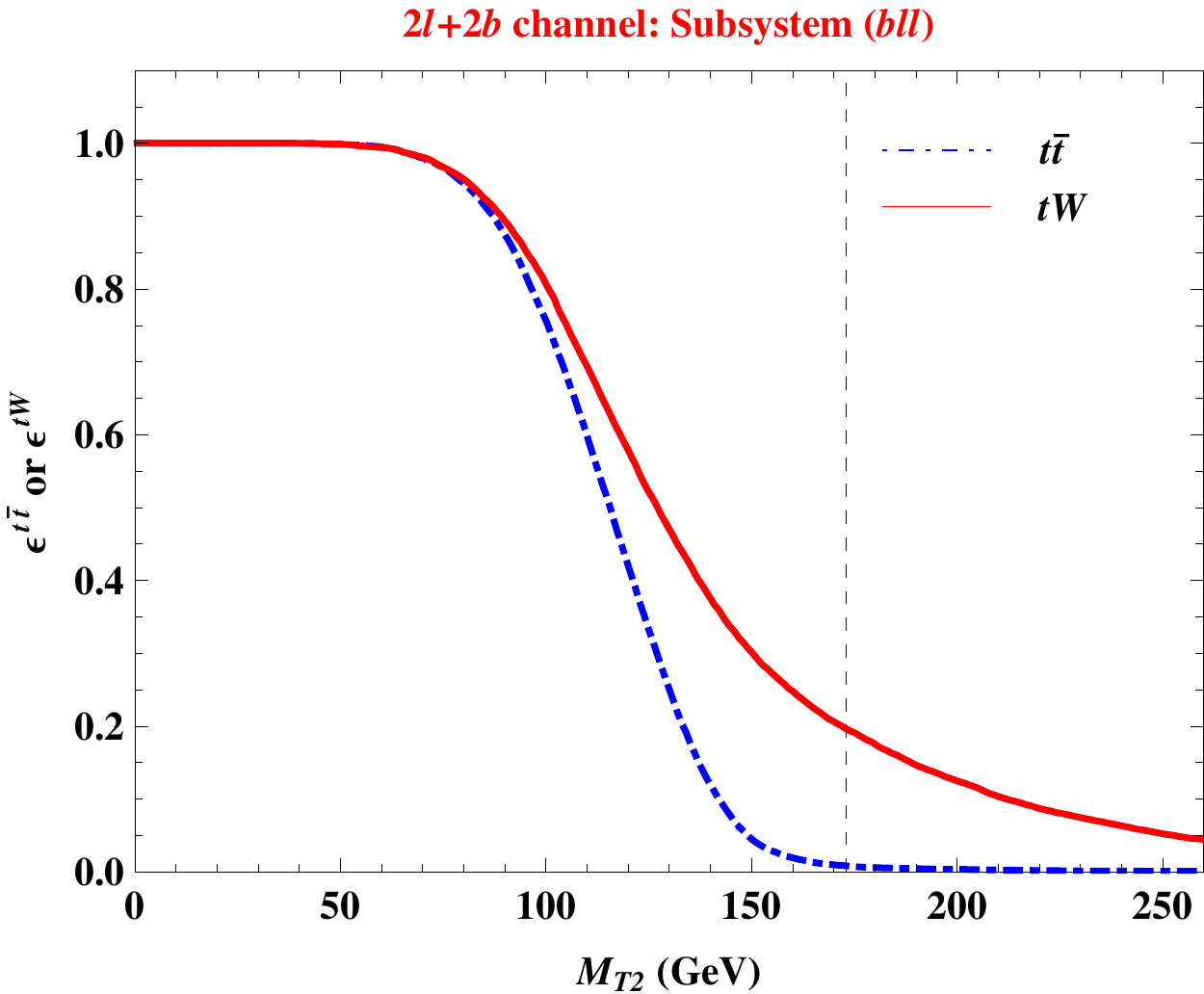}   }
\vspace{0.1cm}
\centerline{
\includegraphics[scale=0.56]{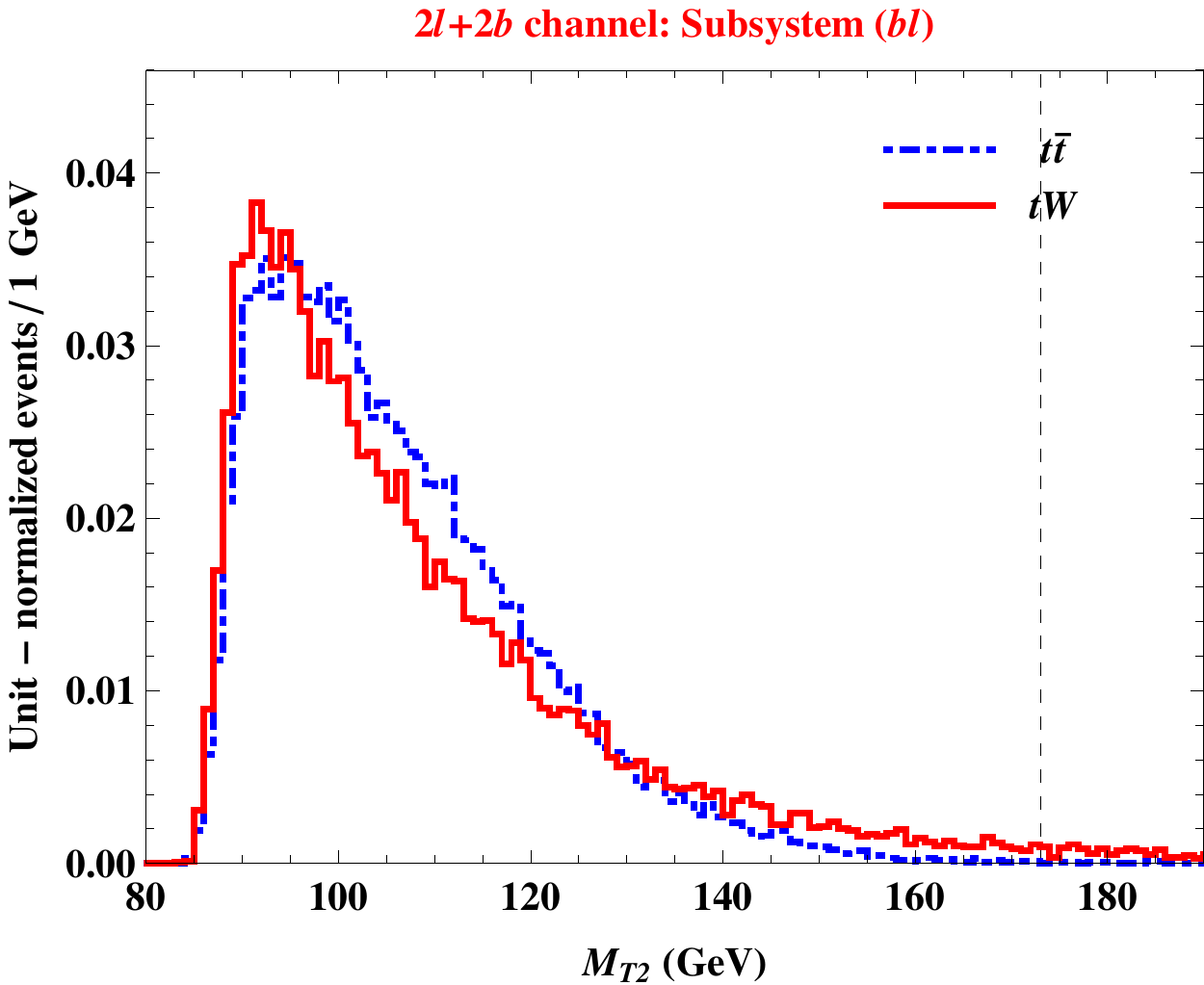}
\hspace{0.4cm}
\includegraphics[scale=0.55]{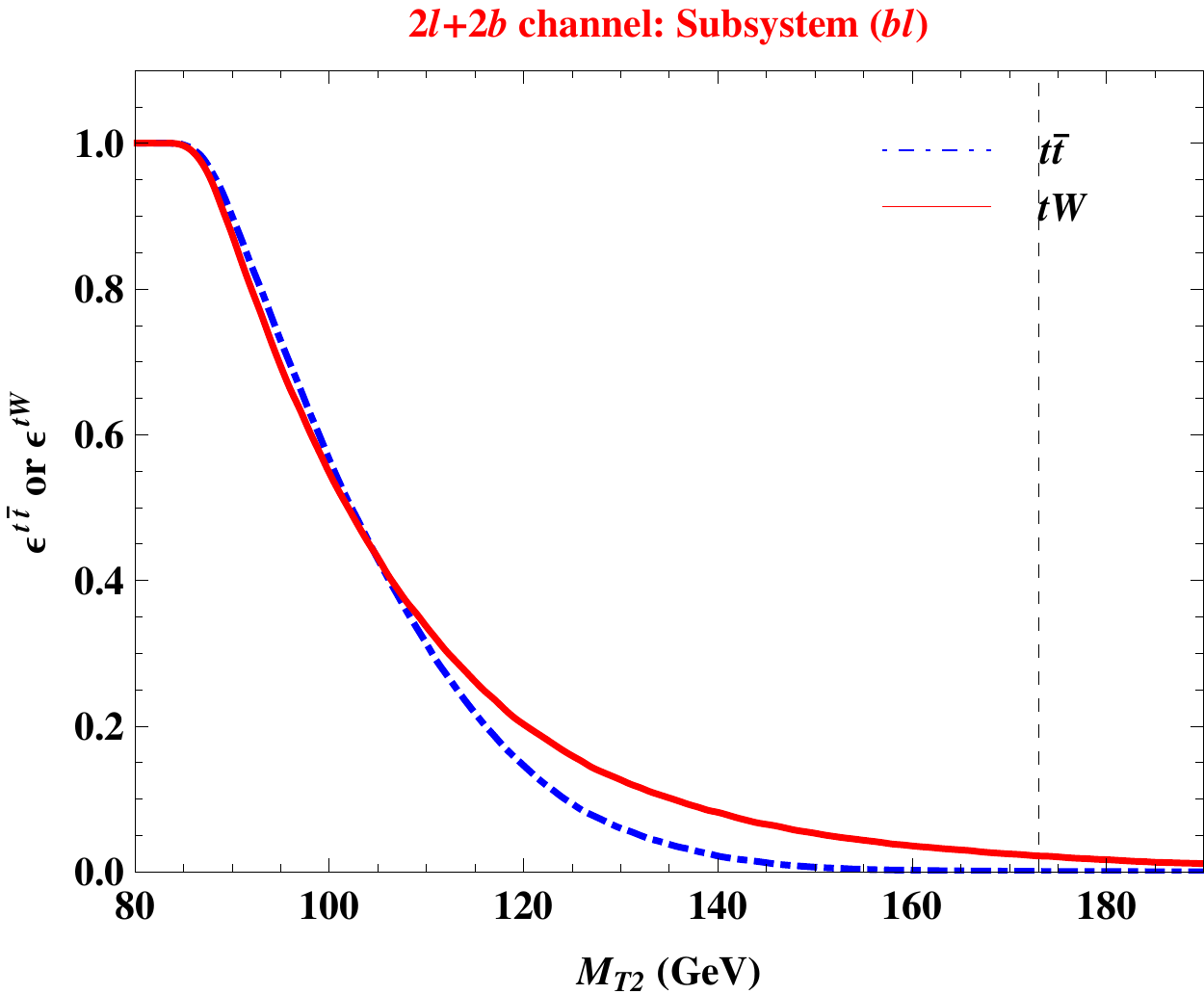}  }
\caption{\label{fig:asym2l2b} $M_{T2}$ distributions of $t\bar{t}$ and $tW$ events for the $(b\ell b)$ (upper-left panel), $(b\ell\ell)$ (middle-left panel), and $(b\ell)$ (lower-left panel) subsystems and
their corresponding selection efficiencies (right panels) in SR-I. The distributions are plotted with the events passing the selection criteria in Eqs.~(\ref{eq:cutfirst}) through~(\ref{eq:cutlast}) with one more $b$-tagged jet is required. The relevant combinatorics arising in all subsystem is treated by the scheme in the text and Table~\ref{tab:selTab}. The test mass for the decay side involving a lepton (only a bottom quark) is 0 GeV (80 GeV). The dashed lines indicate the expected endpoints of the $t\bar{t}$ system. }
\end{figure*}

A similar analysis can be conducted for the three asymmetric subsystems that are exhibited in Figure~\ref{fig:asym2l2b}: the $(b\ell b)$ subsystem in the top panels, the $(b\ell\ell)$ subsystem in the middle panels, and the $(b\ell)$ subsystem in the bottom panels. The $M_{T2}$ distributions are shown in the left panels, while the corresponding efficiency plots are shown in the right panels. Since the invisible particles in the $(b\ell b)$ and $(b\ell)$ subsystems are different in both decay legs, the relevant test masses are applied accordingly, i.e., 0 GeV for the decay leg involving a lepton and 80 GeV for the decay leg involving only a bottom. On the other hand, the $(b\ell\ell)$ subsystem assumes identical invisible particles (here neutrino) so that a common test mass of 0 GeV is employed. Note that there arises a combinatorial issue for all asymmetric subsystems. For any given event, there are two partitionings depending on the way of grouping one lepton and one bottom quark, and for each partitioning two $M_{T2}$ values are available. To resolve this combinatorial ambiguity, we follow the prescription used in Ref.~\cite{Chatrchyan:2013boa} with a slight modification, summarizing as follows. As mentioned above, each partitioning has two $M_{T2}$ values, smaller and larger. Suppose that for one partitioning we have the smaller value $a$ and the larger value $A$, while for the other partitioning we have the smaller value $b$ and the larger value $B$. When ordering those four values, we have six possibilities. As one of the partitionings is correct, either $A$ or $B$ is surely correct. However, we are unaware {\it a priori} which is the case. Here we simply choose the smaller out of $A$ and $B$ as a conservative approach. For the $tW$ with an extra jet, this prescription is subtle because the relevant kinematic endpoint can be arbitrarily high as explained before. But we apply this selection scheme for every single event as if it belonged to the dileptonic $t\bar{t}$. Those ordering and selection rule are tabulated in Table~\ref{tab:selTab}. In principle, this selection scheme is not unique, and other possibilities are still available (see Ref.~\cite{Baringer:2011nh}, for example. Ref.~\cite{Baringer:2011nh} also investigated efficiencies and purities by varying invariant mass and $M_{T2}$ cuts). We attempted other possible selection schemes and found that the above-described prescription is the best for signal-background separation.	

\begin{table}[t]
\centering
\begin{tabular}{c|c c c c c c}
Ordering & $bBaA$ & $aAbB$ & $baBA$ & $baAB$ & $abBA$ & $abAB$ \\
\hline
Selection & $B$ & $A$ & $B$ & $A$ & $B$ & $A$
\end{tabular}
\caption{\label{tab:selTab} Six possible orderings in $m_{b\ell}$ and $M_{T2}$ of three asymmetric subsystems and selection scheme in each ordering. For each ordering, the left-to-right sequence is from the lowest value to the highest. Out of four values, only the values in the second row are plotted in the relevant distributions.}
\end{table}    

Producing the distributions in Figure~\ref{fig:asym2l2b} according to the prescription, we observe a clear separation between the signal and background events in all three subsystems. Most of the background events are populated below the expected kinematic endpoint for the $t\bar{t}$ while a large number of signal events can be found even beyond the endpoint. Again, if the cut is applied near the kinematic endpoint, most of the background events can be suppressed with many signal events kept. This expectation is consistently supported by the associated efficiency curves in the $M_{T2}$ cuts. Like the cases in the symmetric subsystems, they also show that the signal efficiency denoted by red solid curves predominates the background efficiency denoted by blue dashed curves as the cuts are near or beyond the $t\bar{t}$ kinematic endpoints indicated by black dashed lines. 

It is interesting to understand this overflow phenomenon of the signal in the $M_{T2}$ distributions of various subsystems by investigating its asymptotic behavior in the presence of a very hard $b$-jet that typically emerges due to a mis-tag of an ISR jet. By definition of $M_{T2}$ given in Eq.~(\ref{eq:MT2def}), it is sufficient to evaluate the global minimum of the transverse mass for the decay side having such a hard $b$-jet, assuming that it is $M_T^{(1)}$ solely for convenience.
\bea
\left(M_T^{(1)}\right)^2 &=&\left(m_T^{v(1)}\right)^2+\tilde{m}_1^2 
      +2\left( E_T^{v(1)}E_T^{q(1)}-\vec{p}_T^{\,v(1)}\cdot \vec{q}_{T}^{\, (1)} \right)  , \nonumber
\eea
where $E_T^{v(1)}$ and $m_T^{v(1)}$ are the transverse energy and transverse mass formed by all visible particles belonging to the first decay side. One then can prove that the global minimum of the above transverse mass is given by
\bea
\left(M_T^{(1)}\right)_{\min}=m^{v(1)}+\tilde{m}_1 \label{eq:minMT} \, ,
\eea
where $m^{v(1)}$ simply implies the invariant mass formed by the relevant visible particles \cite{Barr:2003rg,Agashe:2010tu}. More specifically, if $m^{v(1)}$ is formed by a bottom and a lepton, it is evaluated by
\bea
\left(m^{v(1)} \right)^2=2E_bE_{\ell}(1-\cos\theta_{b\ell}) \label{eq:mv1} \, ,
\eea
where $\theta_{b\ell}$ denotes the intersecting angle between $b$ and $\ell$. One can easily see that it can be arbitrarily large as the bottom becomes arbitrarily hard unless $b$ and $\ell$ are extremely collinear. Thus, Eq.~(\ref{eq:minMT}) can be arbitrarily large, and in turn, so can $M_{T2}$. This argument is readily applicable to the subsystems where at least one of the decay sides involves a lepton and a bottom at the same time: for example, subsystems $(b\ell b\ell)$, $(b\ell b)$, and $(b\ell\ell)$.

If $m^{v(1)}$ vanishes, however, this argument gets subtle, and thus it is better to look at the full expressions of both $M_T$'s:  

\begin{widetext}
\begin{eqnarray}
\left(M_T^{(1)}\right)^2&=&\tilde{m}_1^2+2\left( p_T^{v(1)}E_T^{q(1)}-\vec{p}_T^{\, v(1)}\cdot \vec{q}_{T}^{\, (1)} \right) \label{eq:MTside1} \, , \\
\left(M_T^{(2)}\right)^2&=&\left(m_T^{v(2)}\right)^2+\tilde{m}_2^2+2\left( E_T^{v(2)}E_T^{q(2)}+\vec{p}_T^{\, v(2)}\cdot (\vec{q}_{T}^{\,(1)}+\vec{p}_T^{\,v(1)}+\vec{p}_T^{\, v(2)}) \right) \nonumber \\
&\approx&\left(m_T^{v(2)}\right)^2+\tilde{m}_2^2+2\left( E_T^{v(2)}E_T^{q(2)}+\vec{p}_T^{\,v(2)}\cdot (\vec{q}_{T}^{\,(1)}+\vec{p}_T^{\,v(1)}) \right) \label{eq:MTside2} \, ,
\end{eqnarray}
\end{widetext}
where in the second line of Eq.~(\ref{eq:MTside2}) we used the assumption that $\vec{p}_T^{\,v(1)}$ is hard enough to dominate over the total visible momentum, i.e., $\vec{p}_T^{\,v(1)}\gg \vec{p}_T^{\,v(2)}$. To minimize Eq.~(\ref{eq:MTside1}), $\vec{q}_T^{\,(1)}$ should be either zero or parallel to $\vec{p}_T^{\,v(1)}$. But then Eq.~(\ref{eq:MTside2}) becomes very large unless $\vec{p}_T^{\,v(1)}$ is anti-parallel to $\vec{p}_T^{\,v(2)}$. On the other hand, to minimize Eq.~(\ref{eq:MTside2}), $\vec{q}_T^{\,(1)}$ should be set to be anti-parallel to $\vec{p}_T^{\,v(1)}$, which makes Eq.~(\ref{eq:MTside1}) become very large. So, the solution is likely to happen in a certain intermediate configuration. However, both Eqs.~(\ref{eq:MTside1}) and~(\ref{eq:MTside2}) are quickly rising as $\vec{q}_T^{\,(1)}$ is away from those extreme configurations due to the largeness of $\vec{p}_T^{\,v(1)}$, and therefore, the final $M_{T2}$ value is very likely to be large. 

\begin{figure*}[t]
\centerline{
\includegraphics[scale=0.61]{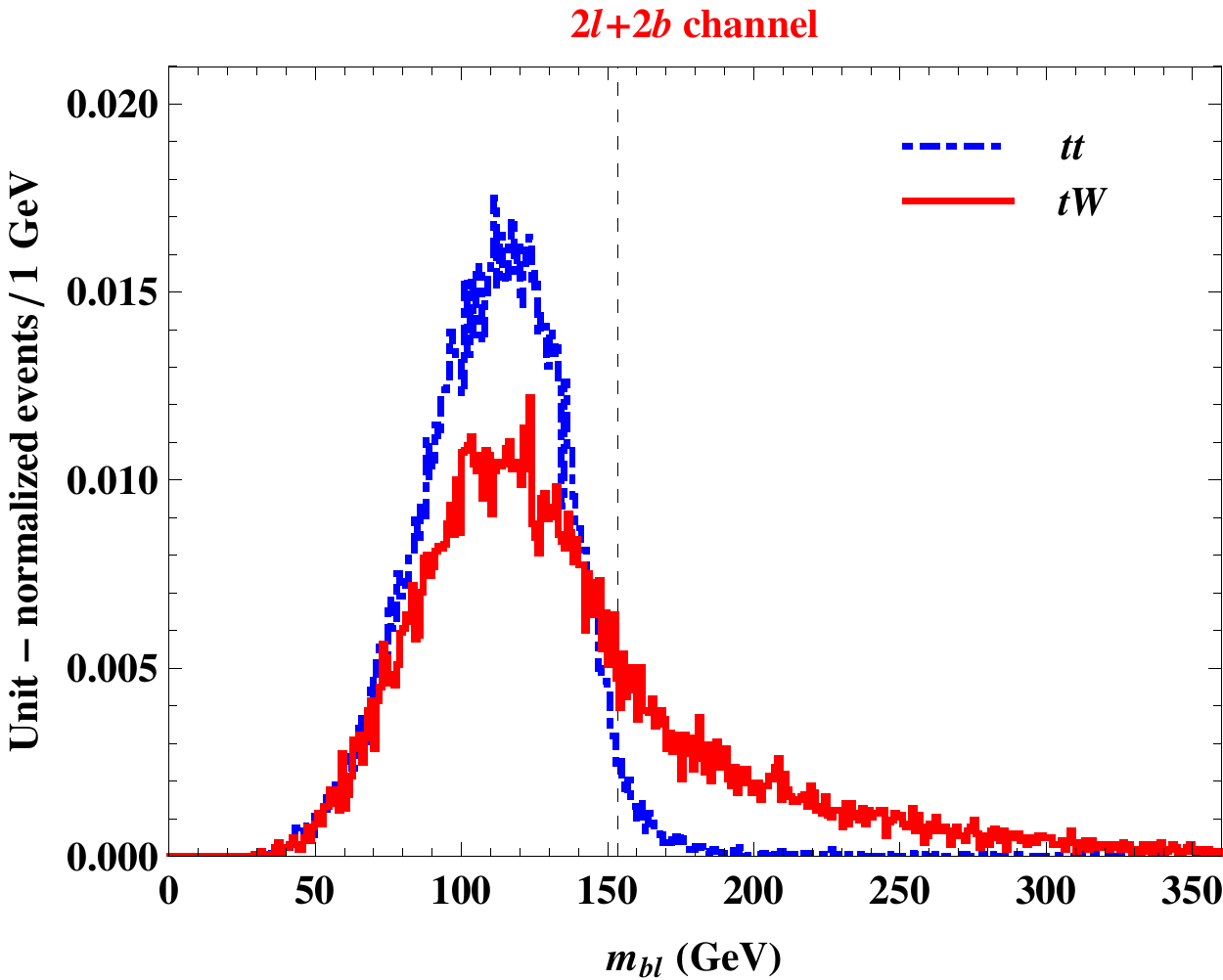}
\hspace{0.4cm}
\includegraphics[scale=0.59]{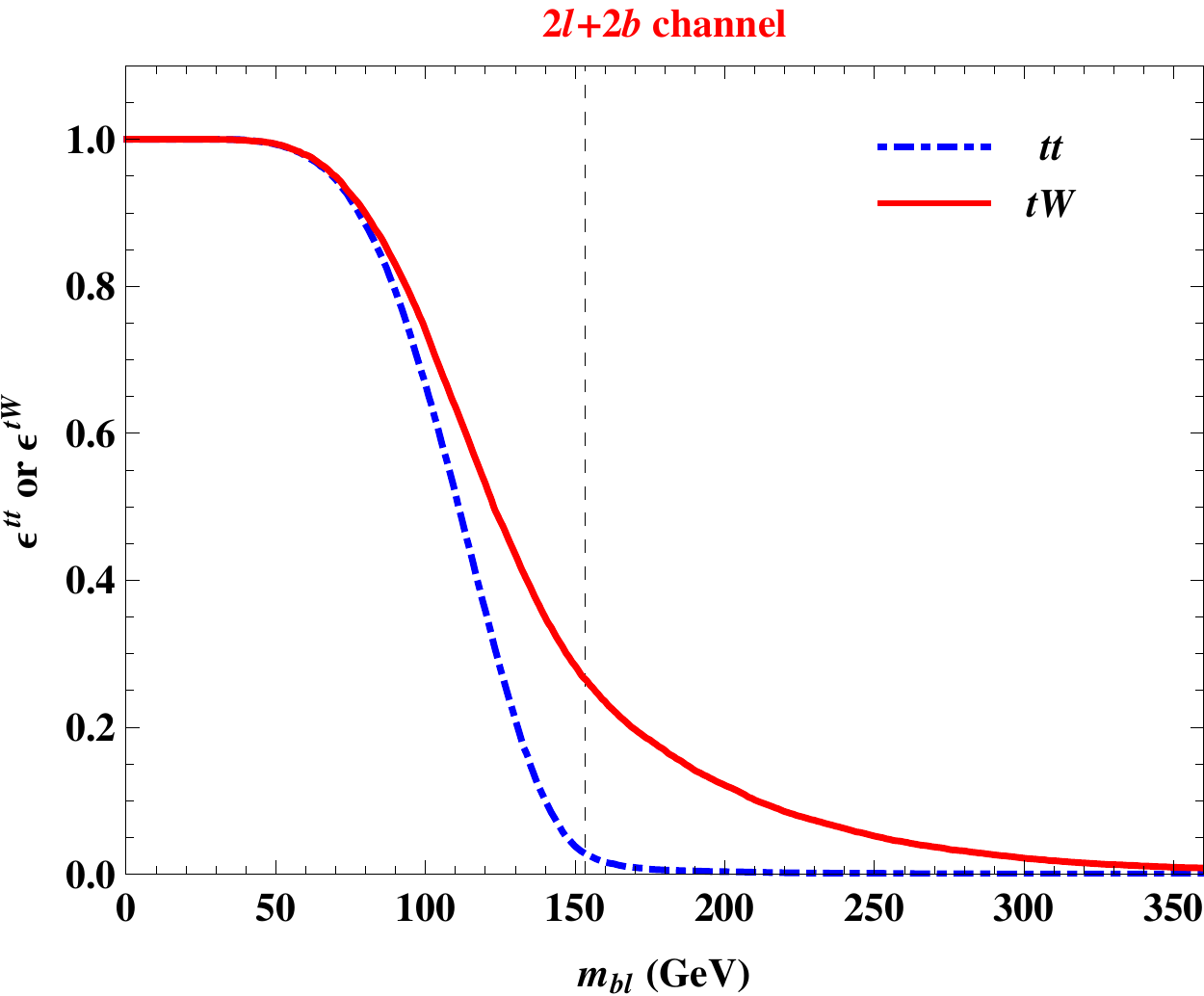} }
\caption{\label{fig:Inv2b2l} Invariant mass distribution (left panel) and $m_{b\ell}$  selection efficiency (right panel) of $t\bar{t}$ and $tW$ events in SR-I. The distributions are plotted with the events passing the selection criteria in Eqs.~(\ref{eq:cutfirst}) through~(\ref{eq:cutlast}) with one more $b$-tagged jet is required. The relevant combinatorics is treated by the prescription explained in the text and Table~\ref{tab:selTab}. The dashed lines indicate the expected endpoints of the $t\bar{t}$ system. }
\end{figure*}
A similar observation can be made for the $m_{b\ell}$ distribution using an analogous argument. Again, the requirement of an additional jet on top of a bottom-tagged jet and two opposite-signed leptons retrieves the entire decay topology of the dileptonic $t\bar{t}$ system, so that the invariant mass variable is upper-bounded as in the case of Sec.~\ref{sec:LO}. On the other hand, the additional jet, which is mis-tagged as a bottom quark in SR-I, can be arbitrarily hard, thus the relevant invariant mass evaluated with it can be arbitrarily large as explained in Eq.~(\ref{eq:mv1}) and thereafter. We therefore expect that the $m_{b\ell}$ distribution for $t\bar{t}$ is bounded above, whereas that for $tW$ is featured by a large tail stretching even beyond the expected $m_{b\ell}$ endpoint of the $t\bar{t}$ system. Obviously, there arises a combinatorial issue in having the $m_{b\ell}$ distributions. For the treatment of wrong combinations in $m_{b\ell}$, we again follow the prescription used in Ref.~\cite{Chatrchyan:2013boa}, being adopted for the $M_{T2}$ variables in the asymmetric subsystems. Having such a selection scheme in our mind, we plot the $m_{b\ell}$ distributions for $t\bar{t}$ and $tW$ in Figure~\ref{fig:Inv2b2l} where the signal and the background distributions are described by the red solid and the blue dashed histograms, respectively. As the selection scheme preserves the kinematic endpoint of the $m_{b\ell}$ distribution for the $t\bar{t}$ system (see also Eq.~(\ref{eq:mblend})), we denote such a theoretical endpoint by the black dashed line. We clearly see that for a large fraction of signal events, the associated $m_{b\ell}$ value exceeds the kinematic endpoint as expected. Like $M_{T2}$, if one imposes a $m_{b\ell}$ cut near the $t\bar{t}$ kinematic endpoint, i.e., keeping the event whose $m_{b\ell}$ value is greater than the cut, one can reject most of the background events while retaining many signal events. The right panel of Figure~\ref{fig:Inv2b2l} shows the associated efficiency curves for the $t\bar{t}$ and $tW$ in the $m_{b\ell}$ cuts. We again observe that the signal efficiency (red solid curve) is better than the background efficiency (blue dashed curve) as the cut is close to or beyond the $t\bar{t}$ kinematic endpoint (black dashed line).

\begin{figure*}[t]
\centerline{
\includegraphics[scale=0.55]{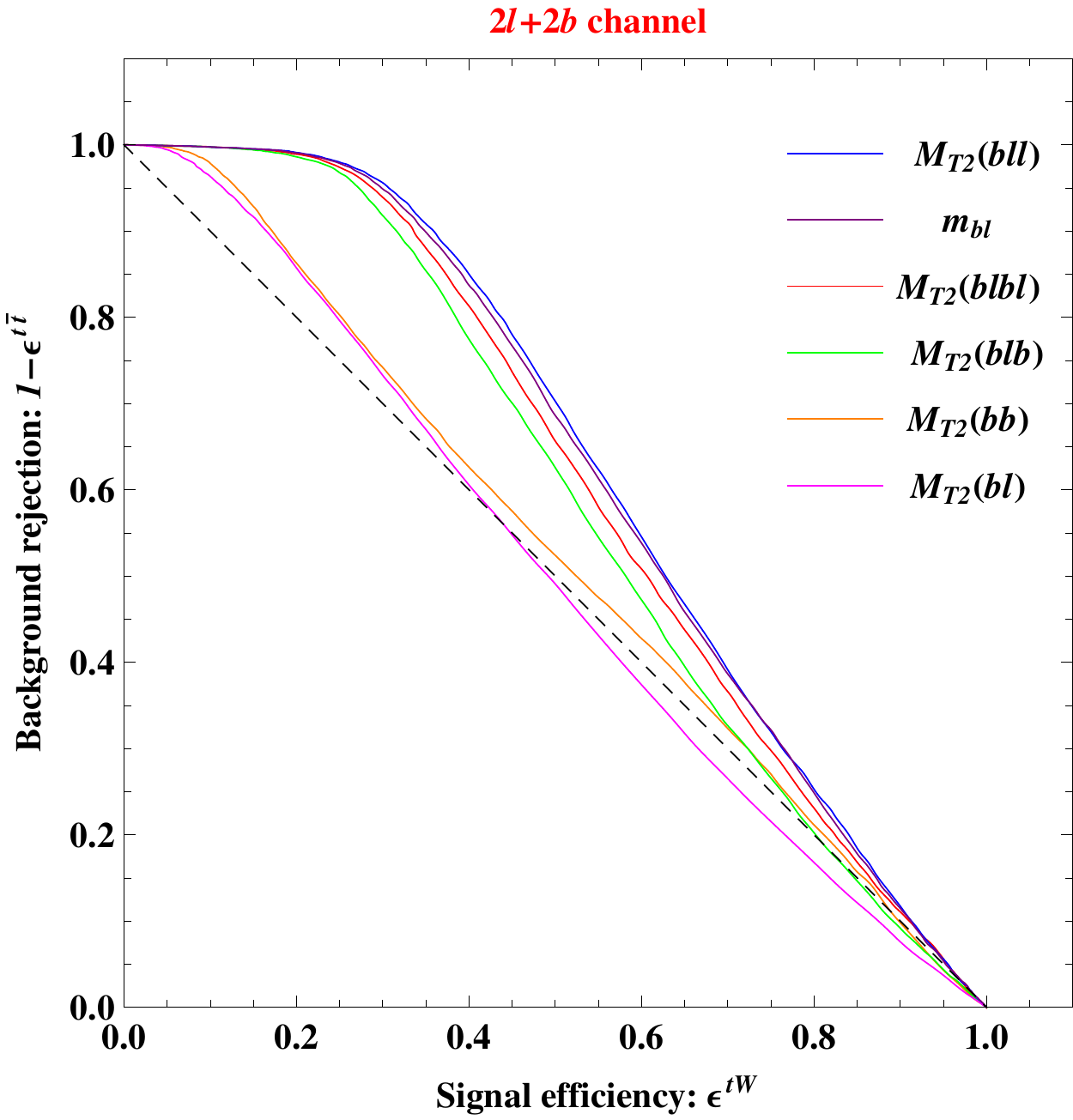}
\hspace{0.8cm}
\includegraphics[scale=0.55]{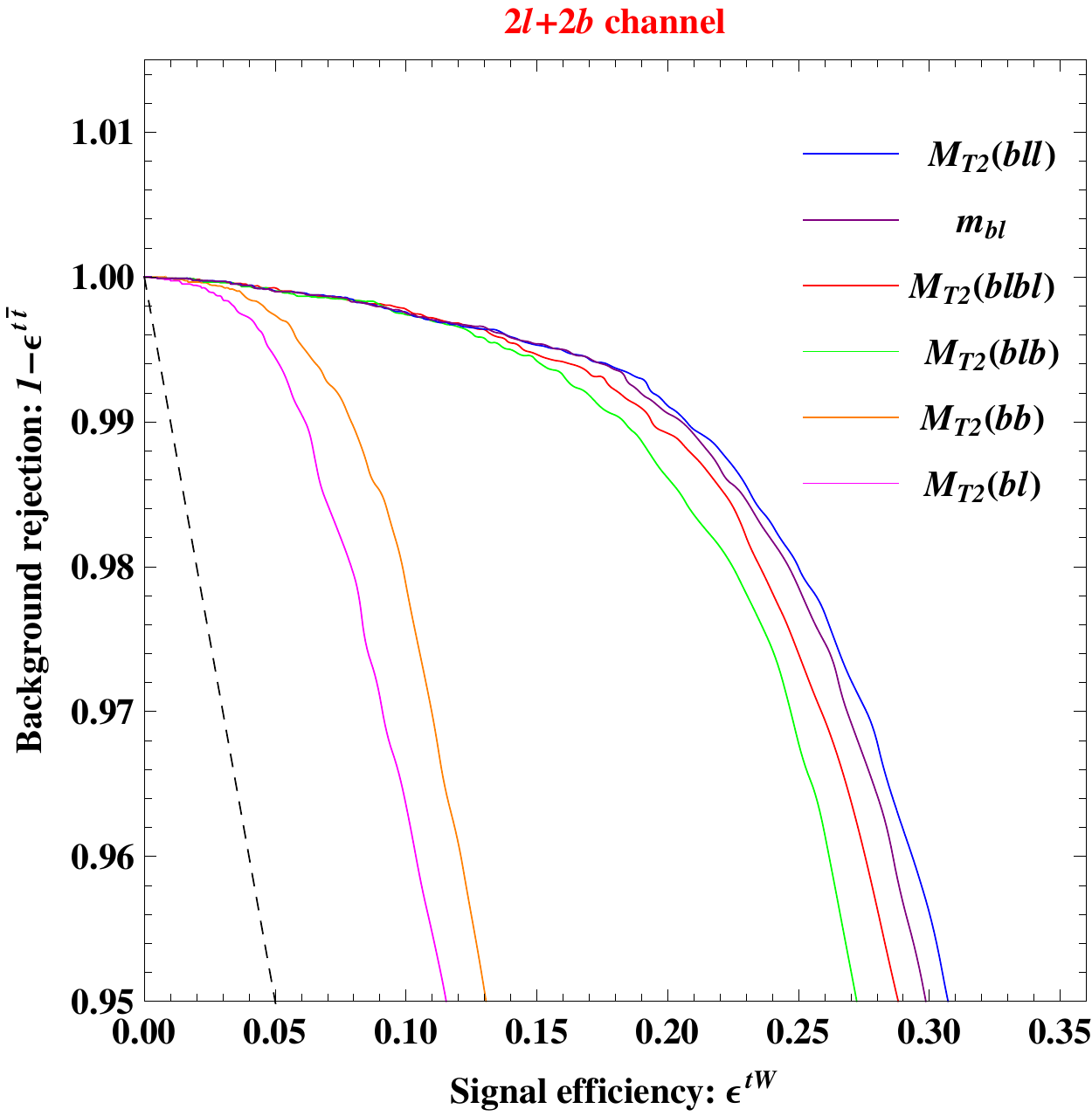} }
\caption{\label{fig:ROC} ROC curves (left panel) for $M_{T2}$ and $m_{b\ell}$ variables and their magnification for the regime having a large background rejection (right panel) in signal region I. }
\end{figure*}

\begin{table*}[t]
\begin{center}
\begin{tabular}{c|c c c c c c}
$1-\epsilon^{t\bar{t}}$ & $M_{T2}(b\ell b\ell)$ & $M_{T2}(bb)$ & $M_{T2}(b\ell b)$ & $M_{T2}(b\ell\ell)$ & $M_{T2}(b\ell)$ & $m_{b\ell}$ \\
\hline \hline
0.999 &	258 (0.056) & 203 (0.036) & 258 (0.052) & 253 (0.050) & 171 (0.024) & 253 (0.049) \\
0.99 & 191 (0.192) & 181 (0.078) & 192 (0.182) & 170 (0.206) & 147 (0.060) & 168 (0.203) \\
0.90 & 164 (0.332) & 159 (0.169) & 167 (0.311) & 143 (0.351) & 125 (0.159) & 140 (0.350)\\
0.50 & 136 (0.601) & 124 (0.522) & 141 (0.579) & 116 (0.623) & 103 (0.475) & 111 (0.626)
\end{tabular}
\caption{\label{tab:ROCtable} Signal efficiency $\epsilon^{tW}$ (numbers in the parentheses) and the associated cuts in GeV for $m_{b\ell}$ and $M_{T2}$ in various subsystems with respect to SR-I. The numbers are tabulated for four representative background rejections, $1-\epsilon^{t\bar{t}}$. }
\end{center}
\end{table*}
To look at the signal-background separation of each variable more closely, we plot the Receiver Operating Characteristic (ROC) curves in Figure~\ref{fig:ROC}. The right panel of it magnifies the region where the background rejections are large. The ROC curve showing the best performance (i.e., large signal efficiency as well as large background rejection) is drawn in the rightmost position, and the others are exhibited in sequence of decreasing performance such as $M_{T2}(b\ell\ell)$, $m_{b\ell}$, $M_{T2}(b\ell b\ell)$, $M_{T2}(b\ell b)$, $M_{T2}(bb)$, and $M_{T2}(b\ell)$. The diagonal line connecting $(1,0)$ and $(0,1)$ (black dotted lines) is drawn for a reference. We here omit the one for the $(\ell\ell)$ subsystem because it is hardly beneficial in selecting signal events against background ones. In other words, it is below or close to the above-mentioned diagonal line in all range. In Table~\ref{tab:ROCtable}, we also tabulate the cuts and signal efficiencies ($\epsilon^{tW}$) of four sample points for which the background events are rejected by a rate of 99.9\%, 99\%, 90\%, and 50\%. The ROC curves suggest that
four variables should provide with almost equally best efficiencies, which are the $m_{b\ell}$ and the $M_{T2}$ in subsystems $(b\ell b\ell)$, $(b\ell b)$, and $(b\ell\ell)$: for example, 99.9\% of background rejection vs. $\sim$5\% of signal acceptance, 99\% of background rejection vs. $\sim$20\% of signal acceptance, and so on. 

\begin{figure*}[t]
\centerline{
\includegraphics[scale=0.56]{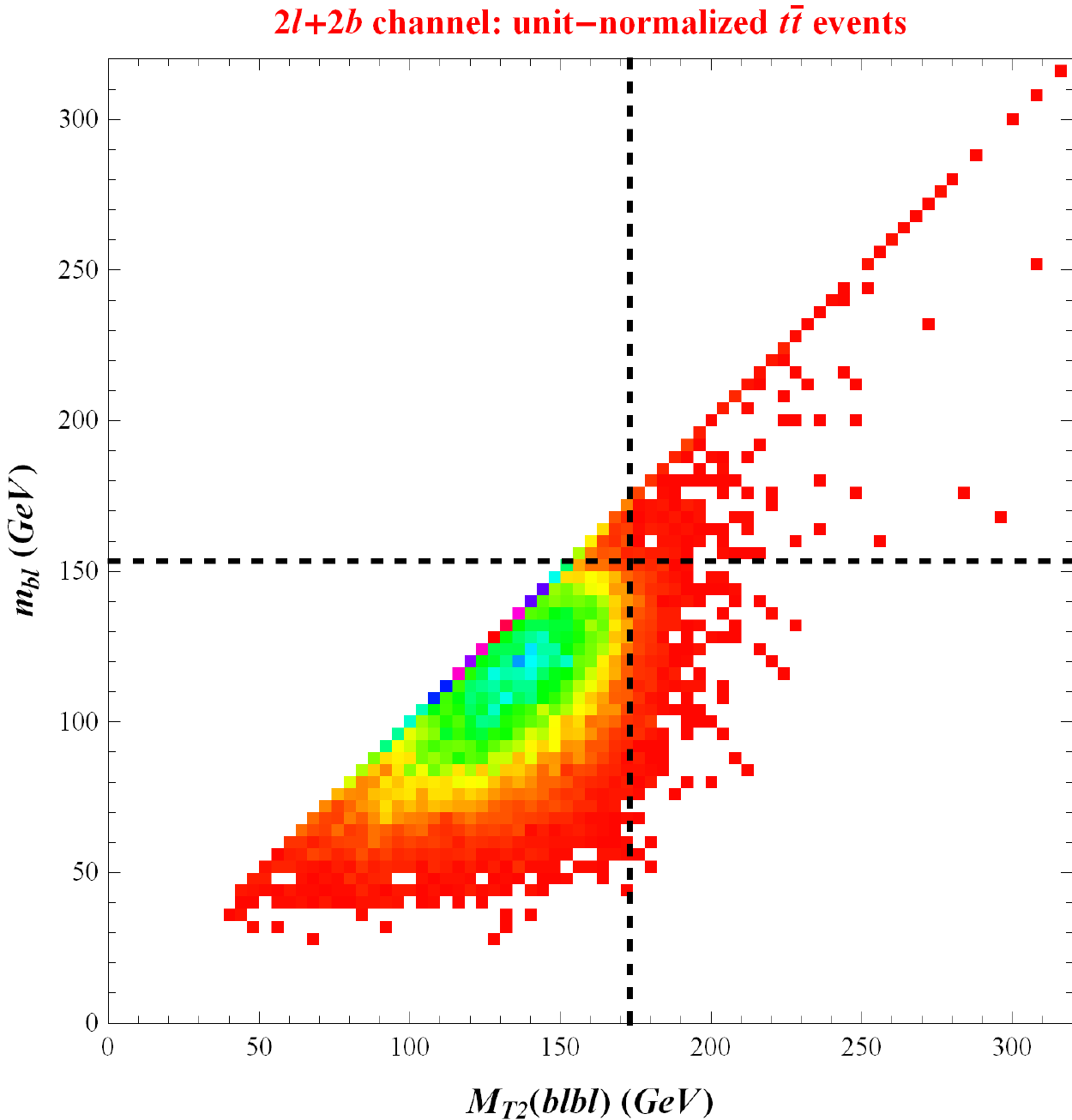}
\hspace{0.8cm}
\includegraphics[scale=0.56]{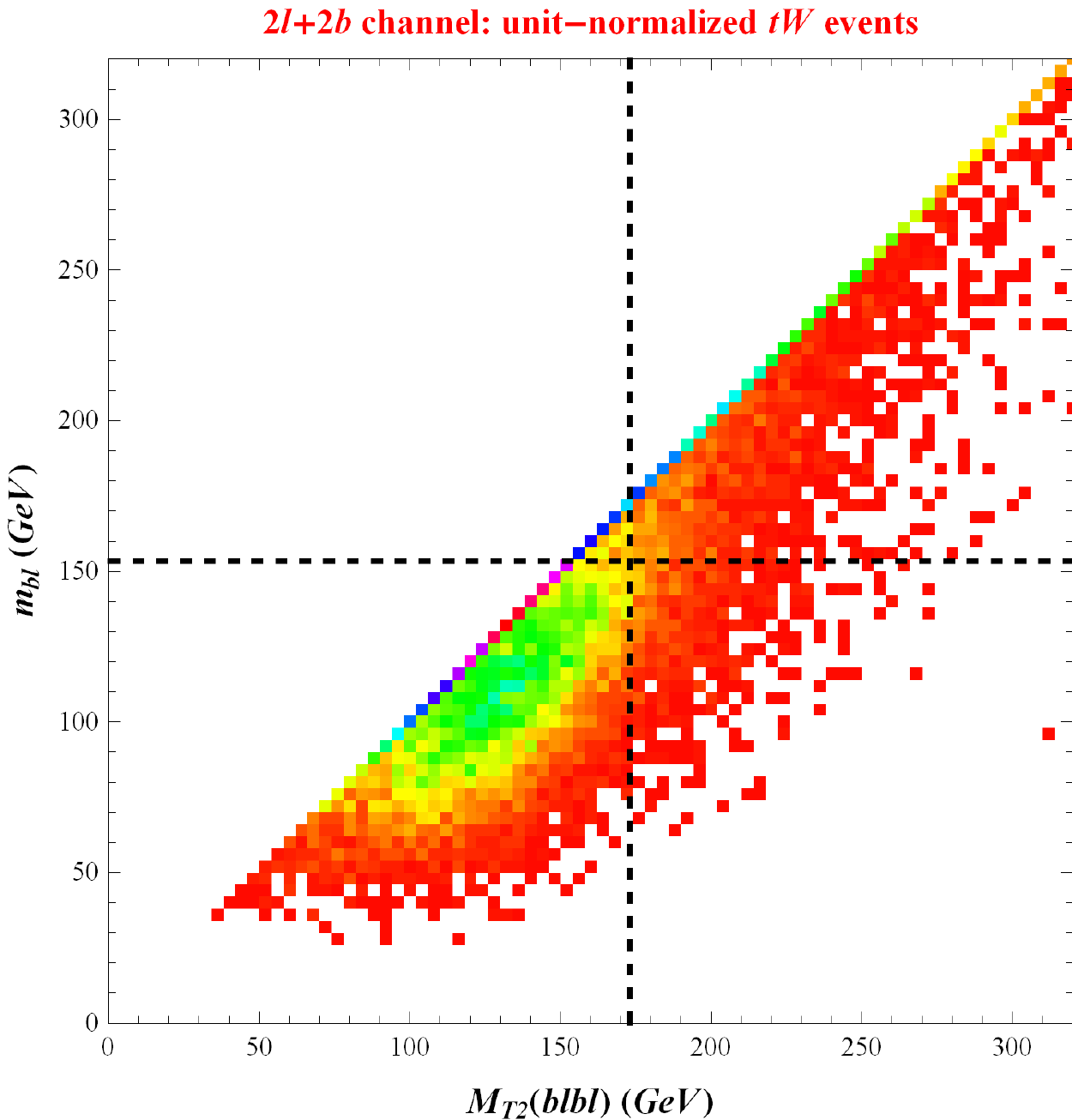}}
\caption{\label{fig:temp} Correlation plots in $M_{T2}(b\ell b\ell)$ vs. $m_{b\ell}$ for $t\bar{t}$ (left panel) and $tW$ (right panel). The vertical and horizontal dashed lines indicate the expected endpoints of $M_{T2}(b\ell b\ell)$ and $m_{b\ell}$ for the $t\bar{t}$ system. }
\end{figure*}

As the above-mentioned four are the best variables, it is interesting to investigate the correlation among them to see if there is any further improvement in the relevant discriminating power. One could attempt various combinations among them. For example, Figure~\ref{fig:temp} demonstrates the unit-normalized two-dimensional temperature plots of $M_{T2}(b\ell b\ell)$ vs. $m_{b\ell}$ for the $t\bar{t}$ (left panel) and the $tW$ (right panel) events. Very roughly, we observe that the two variables have a positive correlation, i.e., as $M_{T2}$ in the $(b\ell b\ell)$ subsystem increases, $m_{b\ell}$ increases as well, and vice versa. In particular, this trend is more manifest for signal events partly because both values are commonly dictated by the hardness of the additional jet. Hence, it is rather challenging to get a dramatic improvement by the introduction of simple schemes such as rejection of events whose $M_{T2}(b\ell b\ell)$ and $m_{b\ell}$ values are simultaneously less than given respective cuts. We instead see that the background events tend to populate in a local region (lower-left corner in the figure), while the signal events spread over a (relatively) wider region. Given this observation, a potential improvement could be achieved by introducing an customized cut enveloping the background region in the left panel of Figure~\ref{fig:temp}. We do not perform a detailed study in this direction because it is beyond the scope of this paper. 

\subsection{Signal region II: $1b+1j+\ell^+\ell^-+\misse$ \label{sec:sr2}}

The same strategy is readily available for Signal Region II. Event selection is done with Eqs.~(\ref{eq:cutfirst}) through~(\ref{eq:cutmiddle}) but a slight modification of Eq.~(\ref{eq:cutlast}) as follows:
\bea
N_j=1 \hbox{ while } N_b=1,\; p_T^{j,b}>30 \hbox{ GeV, } |\eta^{j,b}|<2.4.~~ \label{eq:cutlastmod}  
\eea
Once the jet is selected in this way, it is considered as another $b$-jet throughout the analysis later on. To preclude the inclusion of any extra loose jet, we additionally require that there should be only one jet even satisfying $p_T^j>20$ GeV and $|\eta^j|<4.9$. Although most of events come from either $tW$ or $t\bar{t}$, SR-II is contrasted with SR-I by a couple of qualitative differences. First, an enhanced signal-over-background is anticipated.   Since the additional jet is typically originated from ISR/FSR gluons, more $tW+j$ events can pass the relevant selection criteria than those in SR-I. On the contrary, the ordinary dileptonic $t\bar{t}$ comes with two bottom quarks at the parton level, so that the requirement of a single regular jet and a single bottom jet reduces the background acceptance by the missing rate of bottom quarks. At the expense of gaining more signal acceptance, the signal separation from the background events becomes less efficient. The reason is that for $t\bar{t}$ there is more possibility that such an extra jet is from ISR which would have been rejected by an additional $b$-tagged jet. Like the signal process $tW+j$, the hard ISR jet can render even $t\bar{t}$ events exceed the expected kinematic endpoint, and as a result, the signal efficiency becomes (slightly) reduced for a given background rejection. 

\begin{figure*}[t]
\centerline{\includegraphics[scale=0.6]{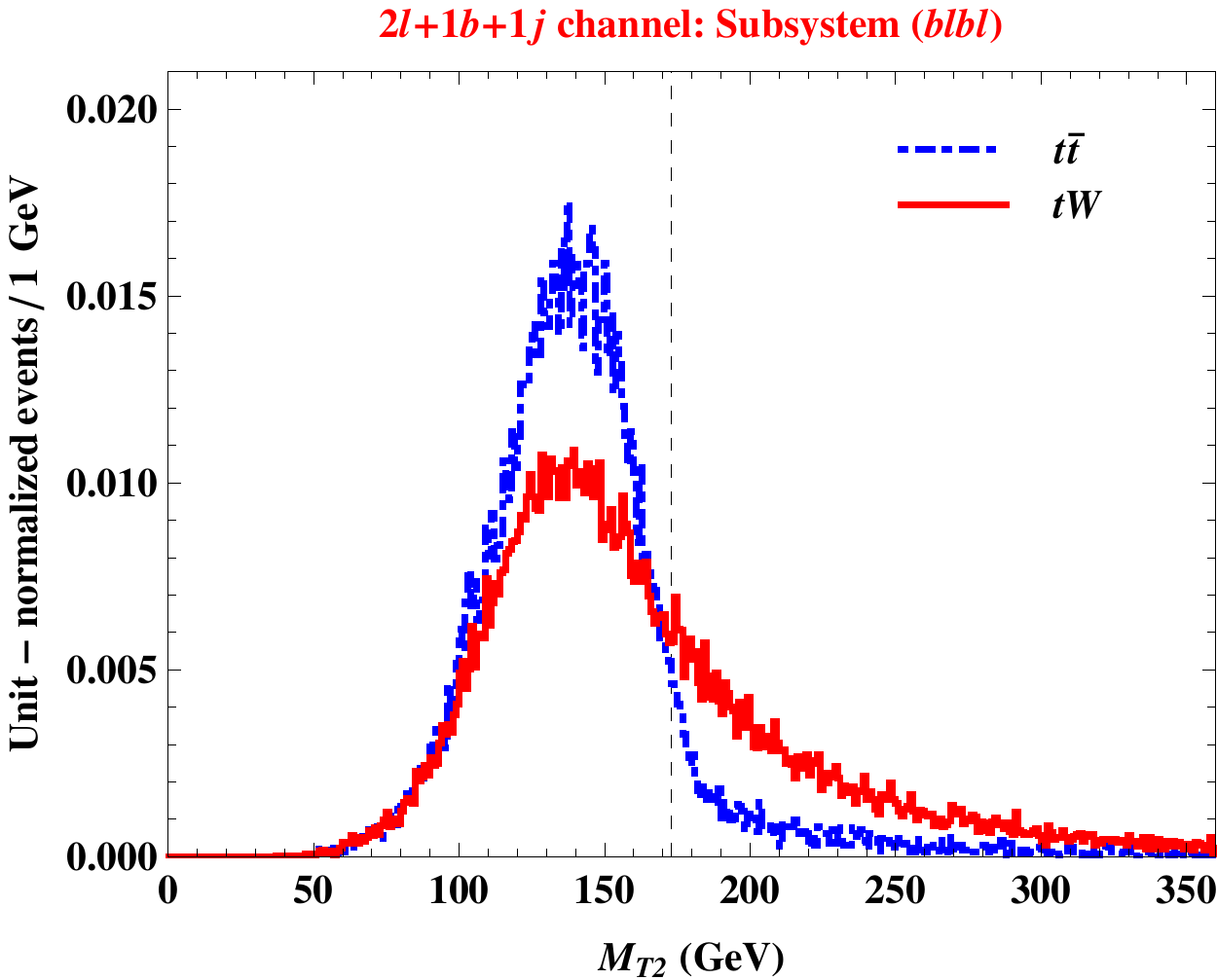} \hspace{0.6cm}
\includegraphics[scale=0.6]{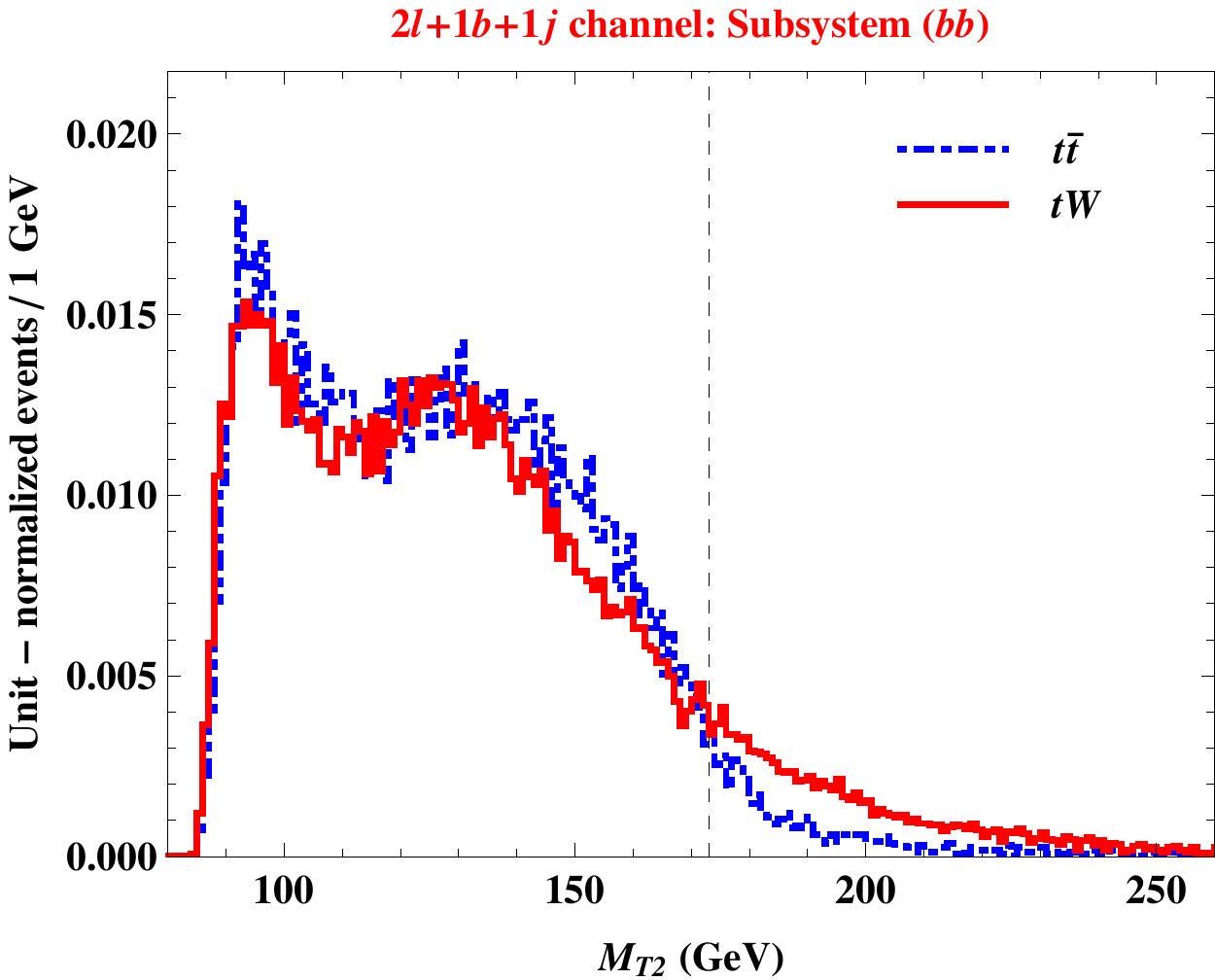} }
\vspace{0.1cm}
\centerline{\includegraphics[scale=0.6]{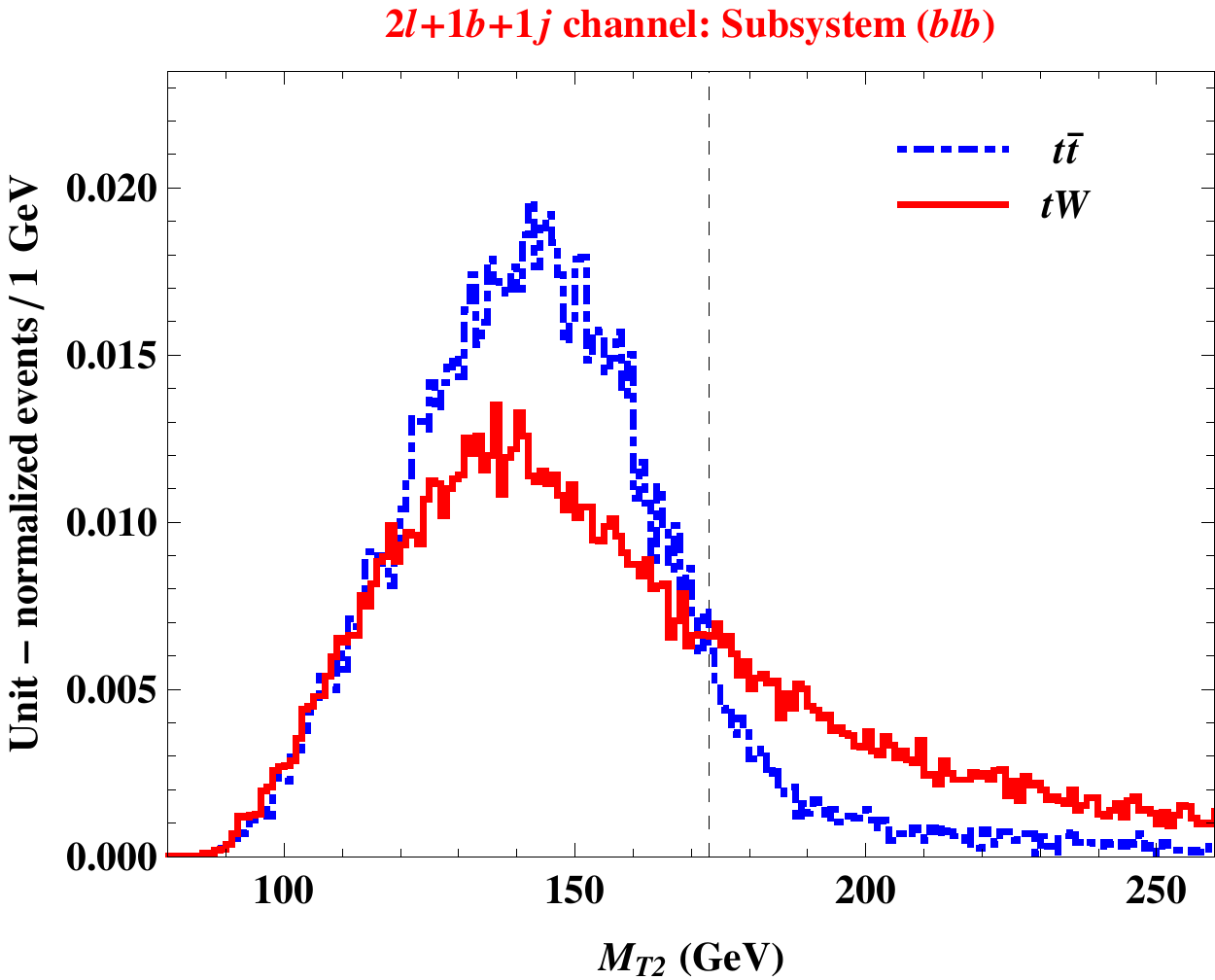} \hspace{0.6cm}
\includegraphics[scale=0.6]{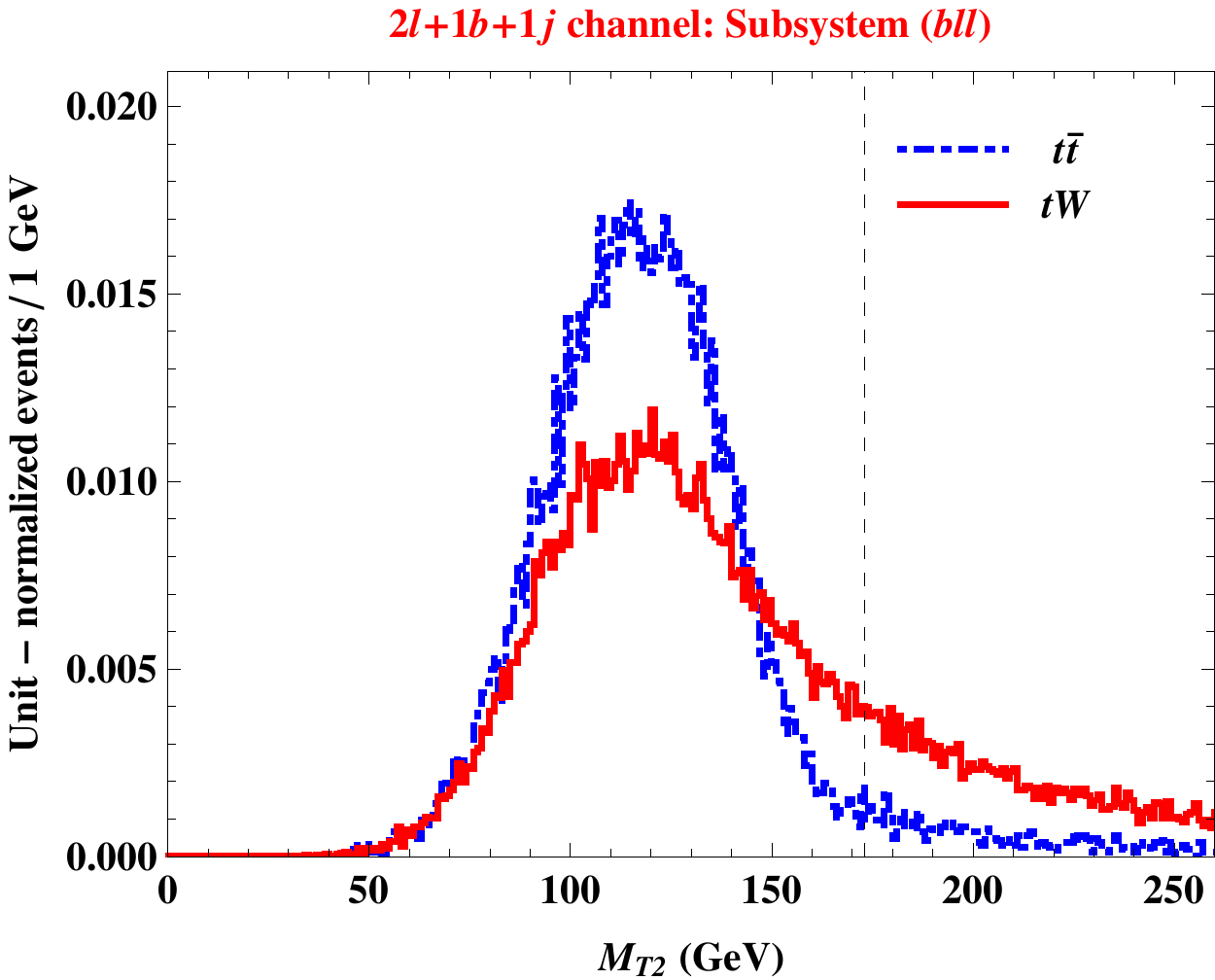}}
\vspace{0.1cm}
\centerline{\includegraphics[scale=0.6]{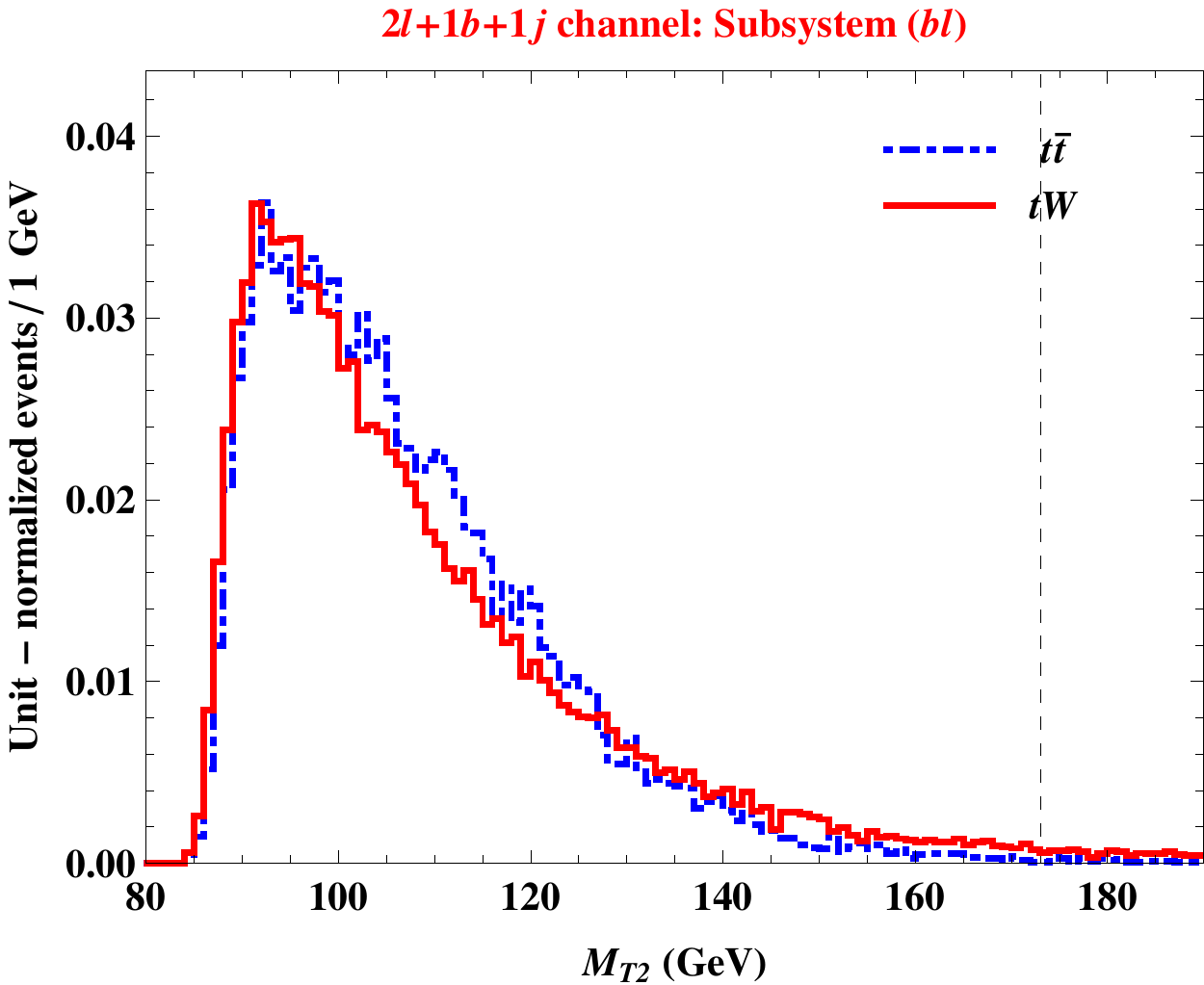} \hspace{0.6cm}
\includegraphics[scale=0.6]{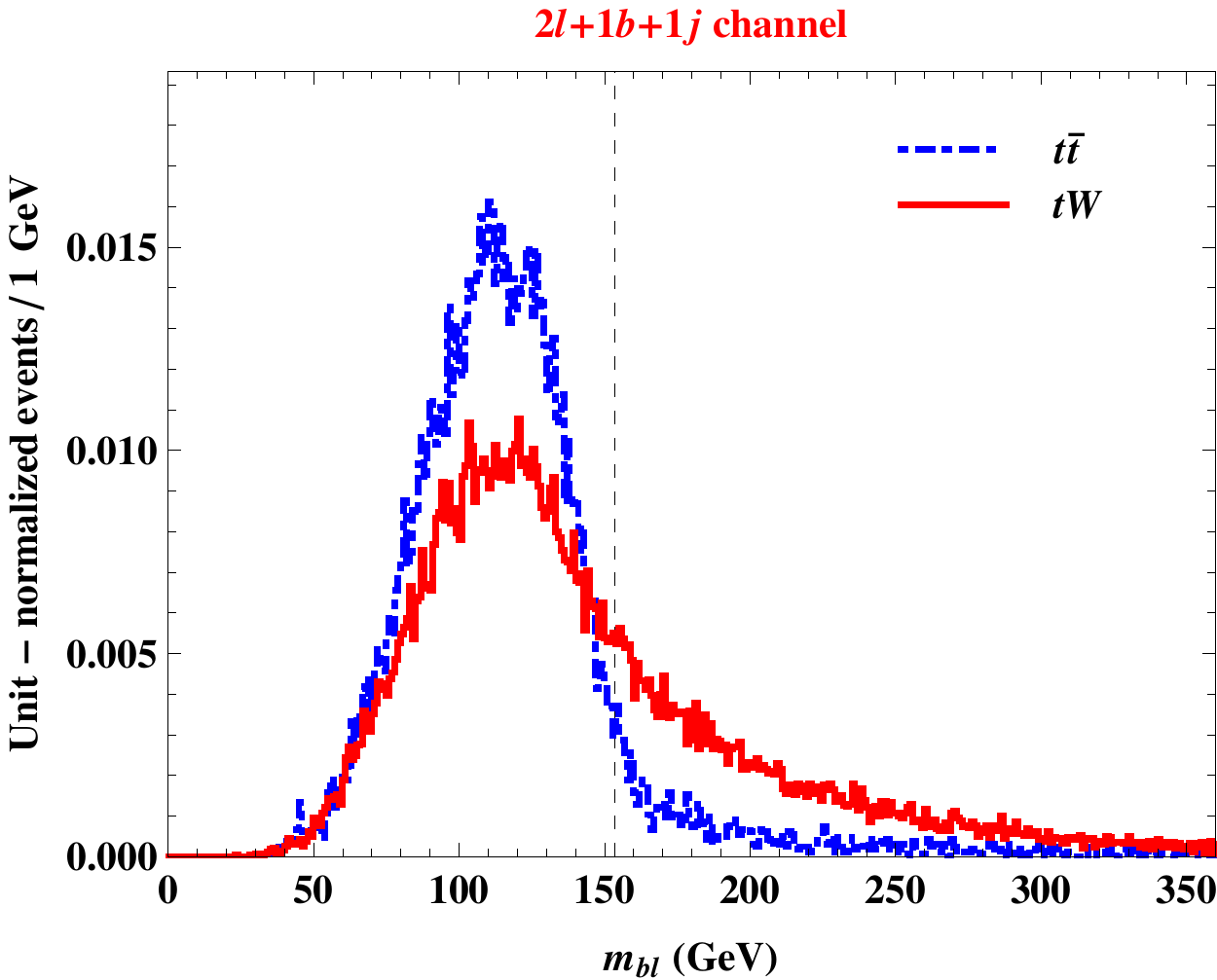} }
\vspace*{-0.3cm}
\caption{\label{fig:2l1b1j} $M_{T2}$ distributions of $t\bar{t}$ and $tW$ events for the $(b\ell b\ell)$ (upper-left panel), $(bb)$ (upper-right panel),  $(b\ell b)$ (middle-left panel),  $(b\ell \ell)$ (middle-right panel), and $(b\ell)$ (lower-left panel) subsystems and $m_{b\ell}$ distribution (lower-right panel) in SR-II. The distributions are plotted with the events passing the selection criteria in Eqs.~(\ref{eq:cutfirst})-(\ref{eq:cutmiddle}) and~(\ref{eq:cutlastmod}). The combinatorics arising in the relevant variables is treated by the scheme in the text and Table~\ref{tab:selTab}. The test mass for the decay side involving a lepton (only a bottom quark) is 0 GeV (80 GeV). The dashed lines indicate the expected endpoints of the $t\bar{t}$ system. }
\end{figure*}

\begin{figure*}[t]
\centerline{
\includegraphics[scale=0.56]{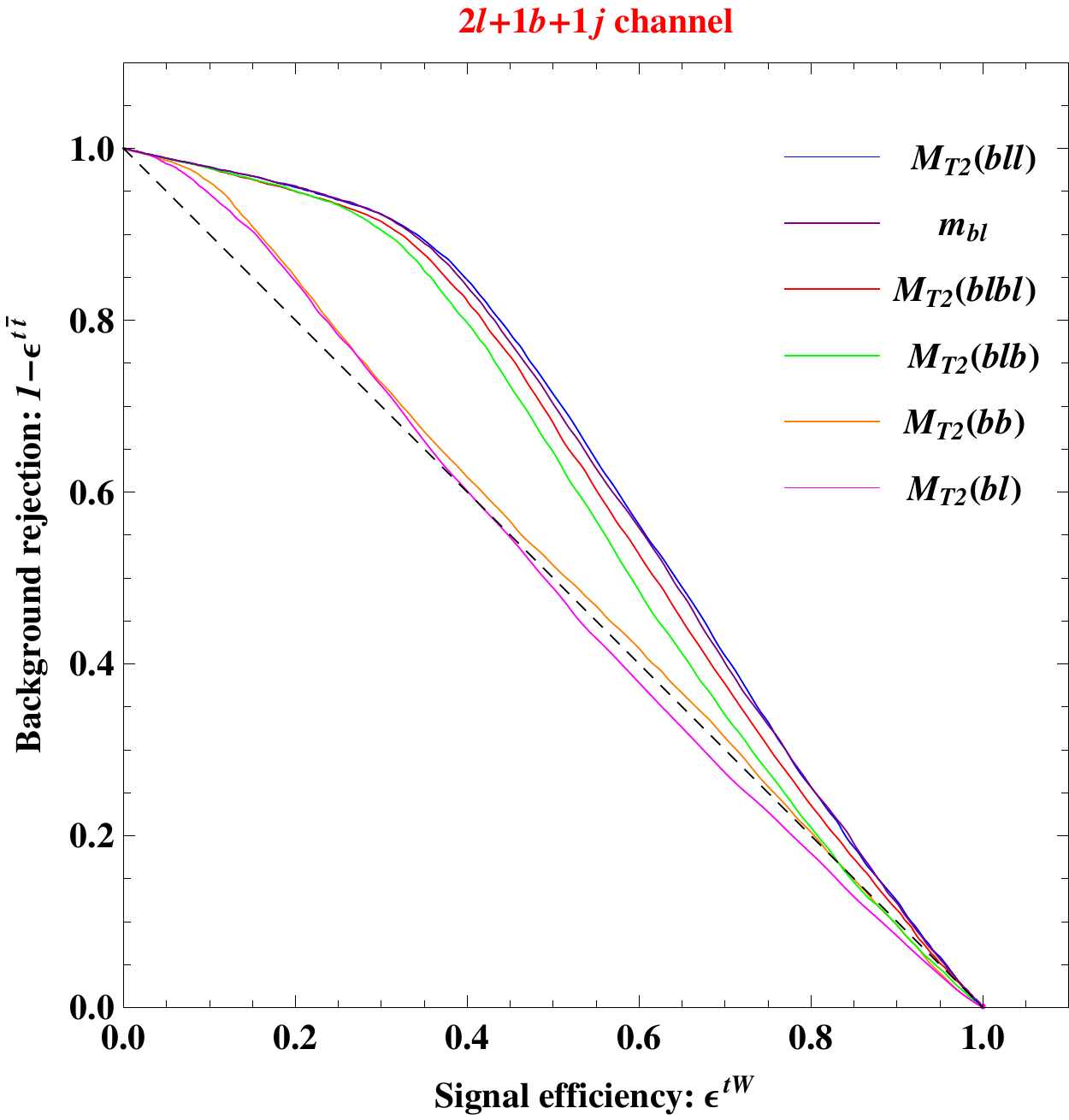}
\hspace{0.8cm}
\includegraphics[scale=0.56]{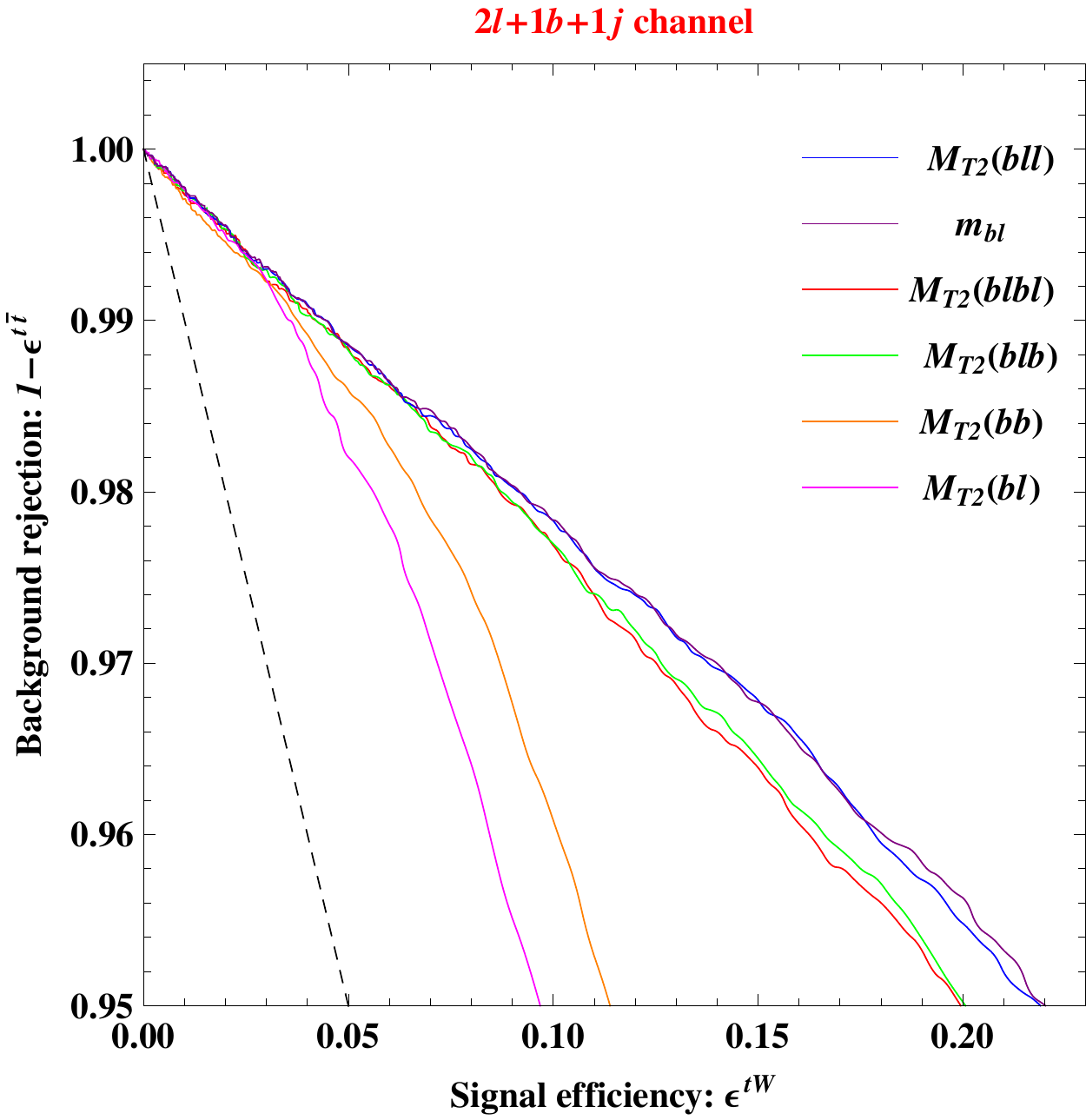} }
\caption{\label{fig:ROCII} ROC curves (left panel) for $M_{T2}$ and $m_{b\ell}$ variables and their magnification for the regime having a large background rejection (right panel) in signal region II. }
\end{figure*}

Figure~\ref{fig:2l1b1j} shows $M_{T2}$ distributions of $t\bar{t}$ and $tW$ events for the $(b\ell b\ell)$ (upper-left panel), $(bb)$ (upper-right panel),  $(b\ell b)$ (middle-left panel),  $(b\ell \ell)$ (middle-right panel), and $(b\ell)$ (lower-left panel) subsystems and $m_{b\ell}$ distribution (lower-right panel). We produce those distributions using the events satisfying the selection criteria given in Eqs.~(\ref{eq:cutfirst})-(\ref{eq:cutmiddle}) and~(\ref{eq:cutlastmod}). The combinatorial ambiguity arising in all variables but the $M_{T2}$ for the $(bb)$ subsystem is taken care of by the same prescriptions elaborated in the previous subsection. The employed test masses are the same as the ones used in the corresponding $M_{T2}$ variables in SR-I. As before, the black dashed lines indicate the expected endpoints of the $t\bar{t}$ system. We observe that all distributions look very similar to the corresponding ones demonstrated in Figures~\ref{fig:MT2blb},~\ref{fig:asym2l2b}, and~\ref{fig:Inv2b2l}. However, we also observe that more background events leak beyond the associated kinematic endpoints as discussed before. To see the correlation of signal acceptance vs. background rejection, we plot the ROC curves in Figure~\ref{fig:ROCII}. Like in SR-I, the right panel of it zoom in the region where the background rejections are large. The color code is the same as that in Figure~\ref{fig:ROC}. More quantitatively, we enumerate the cuts and signal efficiencies of four sample points in Table~\ref{tab:ROCtable2l1b1j} like Table~\ref{tab:ROCtable}. Signal acceptance is somewhat worse than that in SR-I for large background rejection. But it becomes improved compared with that in SR-I as background rejection decreases. 

\begin{table*}[t]
\centering
\begin{tabular}{c|c c c c c c}
$1-\epsilon^{t\bar{t}}$ & $M_{T2}(b\ell b\ell)$ & $M_{T2}(bb)$ & $M_{T2}(b\ell b)$ & $M_{T2}(b\ell\ell)$ & $M_{T2}(b\ell)$ & $m_{b\ell}$ \\
\hline \hline
0.999 &	480 (0.003) & 278 (0.003) & 473 (0.003) & 451 (0.005) & 217 (0.004) & 451 (0.004) \\
0.99 & 297 (0.042) & 202 (0.036) & 292 (0.042) & 285 (0.044) & 159 (0.035) & 284 (0.044) \\
0.90 & 174 (0.318) & 162 (0.155) & 175 (0.305) & 152 (0.340) & 126 (0.153) & 148 (0.346)\\
0.50 & 138 (0.617) & 125 (0.513) & 143 (0.587) & 118 (0.635) & 103 (0.485) & 113 (0.635)
\end{tabular}
\caption{\label{tab:ROCtable2l1b1j} Signal efficiency $\epsilon^{tW}$ (numbers in the parentheses) and the associated cuts in GeV for $m_{b\ell}$ and $M_{T2}$ in various subsystems with respect to SR-II. The numbers are tabulated for four representative background rejections, $1-\epsilon^{t\bar{t}}$. }
\end{table*}

\section{\label{sec:discussion} Discussions and outlook}

The top quark is the heaviest particle in the Standard Model and has the largest coupling to the Higgs boson.
It may open up a new window toward new physics and therefore it is important to understand its properties.
Very recently, production of a top quark in association with a $W$ boson has been observed by the ATLAS and CMS collaborations.
Most of kinematic properties of the signal ($tW$) are very similar to those of $t\bar{t}$ that is the dominant background.  
Multi-Variate Analysis has been adapted to discover the production of $tW$ without detailed understanding of kinematics of the signal and its backgrounds. 

In this paper, we have re-examined the production of the single top and a $W$ gauge boson in the Standard Model with a non-conventional strategy.
Our suggestion is to consider $tW + j$ instead of $tW$, which also modifies relevant backgrounds correspondingly. 
This next-to-leading order production for $tW$ signifies the retrieval of the visible state of ordinary $t\bar{t}$, the major background, under the assumption that such an additional jet mostly comes from one of the bottom quarks in it. Clearly, the relevant kinematic structure of the background is well-defined, so that the distributions in well-known kinematic variables such as the invariant mass and $M_{T2}$ are featured by {\it well}-defined kinematic endpoints. This is contrasted with the ill-defined kinematic structure for $tW+j$ due to the fact that $j$ is typically from ISR/FSR. As a consequence, it was observed that for $tW+j$, the kinematic endpoints of aforementioned distributions are also {\it ill}-defined, i.e., the distributions are not bounded above. Based on these observations, we found that one could suppress $t\bar{t}$ background very efficiently with those variables, while obtaining a high efficiency in the signal, 
The simple use of kinematic variables could have helped the earlier discovery by a large significance in combination with conventional channels. 
Since this method provides excellent background rejection, one could try to study other properties of top quark in this channel. We strongly encourage the ATLAS and CMS collaborations to revisit their study on $tW$ with our suggestions. Moreover searches for $B^\prime$ (bottom partner) in the $tW$ final state may exploit the similar techniques.

We emphasize that our novel strategy is very general and can play a key role in separating signal and background events even in the context of physics models beyond the Standard Model. More specifically, the discussion in this paper is readily  applicable to any processes that resemble the following structure:
\begin{eqnarray}
A \bar{A}     &\to& (B b) \,  (\bar{B} \bar{b})    \to (C c \, b)  \,  ( \bar{C} \bar{c} \, \bar{b})  \, , \\
A \Bar{B}   &\to& (B b)\,  ( \bar{B}           )     \to (C c \, b) \,   ( \bar{C} \bar{c} ) \, ,
\end{eqnarray}
where the former represents pair-production of particle $A$ while the latter represents single-production of particle $A$ in association with particle $B$. Here $A \to B b$ ($\bar{A} \to \bar{B} \bar{b}$), $B \to C c$ ($\bar{B} \to \bar{C} \bar{c}$), and the bar denotes anti-particle. In supersymmetric models, one can imagine the following processes.
\begin{itemize}
\item[(1)] $\tilde{t} {\tilde{t}^*}$ vs. $\tilde{t} \tilde{\chi}_1^-$ (or $\tilde{t}^* \tilde{\chi}_1^+$) where 
$\tilde{t} \to \tilde{\chi}_1^+ b \to b\, \ell^+ \tilde{\nu}$ and similarly $\tilde{t}^*  \to \tilde{\chi}_1^- \bar{b} \to \bar{b}\, \ell^- \tilde{\nu} $
\item[(2)] $\tilde{g} {\tilde{g}}$ vs. $\tilde{g} \tilde{q}$ (or $\tilde{g} \tilde{q}^*$) where $\tilde{g} \to q {\tilde{q}} \to q \bar{q}\tilde{\chi}_1^0 $ 
\end{itemize}
The selection procedure targeting at the full visible state of the former processes inevitably demands an extra object for the latter ones, leading an ill-defined event topology for the latter ones only. Then the kinematic variable-based strategy proposed in this paper can help us separate the latter processes from the former ones. 

\section*{Acknowledgments}
We thank Tanumoy Mandal, Stephen Mrenna, and Danny Noonan for useful discussions. 
D. K. is supported by the LHC Theory Initiative postdoctoral fellowship (NSF Grant No.~PHY-0969510) and 
K. K. is supported by the U.S. DOE under Grant No. DE-FG02-12ER41809 and by the University of Kansas General Research Fund allocation 2301566.

\end{document}